\documentclass[preprint,sort&compress,12pt]{elsarticle}
\usepackage{bbm}
\usepackage{amstext}
\usepackage{amsmath}
\usepackage{amssymb}
\usepackage{mathrsfs}
\usepackage{stmaryrd}
\usepackage{bm}
\usepackage{pstricks}
\usepackage{pst-coil}
\usepackage{pst-3d}
\usepackage{amsfonts}
\usepackage{amsthm}
\usepackage{CJK}

\newcommand{\mm}{\mathrm}
\newcommand{\ml}{\mathcal}

\newcommand{\be}{\begin{equation}}
\newcommand{\bea}{\begin{equation}\begin{aligned}}
\newcommand{\beas}{\begin{equation*}\begin{aligned}}
\newcommand{\eeas}{\end{aligned}\end{equation*}}
\newcommand{\eea}{\end{aligned}\end{equation}}
\newcommand{\ee}{\end{equation}}

\renewcommand{\div}{{\rm div }}

\begin{document}
\begin{frontmatter}
\title{
A New Upper Bound for the Largest Growth Rate of \\
Linear Rayleigh--Taylor Instability}

\author[AB]{Changsheng Dou}
\ead{csdou010@163.com}
\author[FJ]{Jialiang Wang}
\ead{2295751521@qq.com}
\author[FJ]{Weiwei Wang}
\ead{wei.wei.84@163.com}
\address[AB]{School of Statistics, Capital University of Economics and Business, Beijing 100070, PR China}
\address[FJ]{College of Mathematics and
Computer Science, Fuzhou University, Fuzhou,   China.}

\begin{abstract}
We investigate the effect of surface tension on the linear Rayleigh--Taylor (RT) instability in stratified incompressible viscous fluids  with or without (interface) surface tension. The existence of linear RT instability solutions with largest growth rate $\Lambda$ is proved under the instability condition (i.e., the surface tension coefficient $\vartheta$ is less than a threshold $\vartheta_{\mm{c}}$) by modified variational method of PDEs. Moreover we find a new upper bound for $\Lambda$. In particular,
we observe from the upper bound that $\Lambda$ decreasingly converges to zero, as $\vartheta$ goes from zero to the threshold $\vartheta_{\mm{c}}$. The convergence behavior of  $\Lambda$ mathematically verifies the  classical RT instability experiment that the instability growth is limited by surface tension during the linear stage.
\end{abstract}
\end{frontmatter}


\newtheorem{thm}{Theorem}[section]
\newtheorem{lem}{Lemma}[section]
\newtheorem{pro}{Proposition}[section]
\newtheorem{concl}{Conclusion}[section]
\newtheorem{cor}{Corollary}[section]
\newproof{pf}{Proof}
\newdefinition{rem}{Remark}[section]
\newtheorem{definition}{Definition}[section]

\section{Introduction}\label{introud}
\numberwithin{equation}{section}

Considering two completely plane-parallel layers of stratified (immiscible) fluids, the heavier one on top of the lighter one and both subject to the earth's gravity, it is well-known that such equilibrium state is unstable to sustain small disturbances, and this unstable disturbance will grow and lead to a release of potential energy, as the heavier fluid moves down under the gravitational force, and the lighter one is displaced upwards. This phenomenon was first studied by Rayleigh \cite{RLIS} and then Taylor \cite{TGTP}, and is called therefore the Rayleigh--Taylor (RT) instability. In the last decades, this phenomenon has been extensively investigated from mathematical, physical and numerical aspects, see \cite{CSHHSCPO,WJH,GYTI1} for instance. It has been also widely investigated how the RT instability evolves under the  effects of other physical factors, such as elasticity \cite{JFJWGCOSdd,FJWGCZXOE,HGJJJWWWOJM,WWWYYZ2018,CYPWWWZYYOE}, rotation \cite{CSHHSCPO,DRJFJSRS}, (internal) surface tension \cite{GYTI2,WYJTIKCT,JJTIWYHCMP}, magnetic fields \cite{JFJSWWWOA,JFJSJMFM,JFJSJMFMOSERT,JFJSWWWN,WYC,WYTIVNMI,JFJSOMITNWZ,JFJSSETEFP} and so on.
In this article, we are interested in the effect of surface tension on the linear RT instability in stratified incompressible viscous fluids. To conveniently introduce relevant mathematical progress and our main results, next we shall mathematically formulate our problem in details.

\subsection{Motion equations in Eulerian coordinates}

Let us first recall a mathematical model, which describes the horizontally periodic motion of stratified  incompressible viscous fluids in an infinity layer domain \cite{JFJWGCOSdd}:
\begin{equation}\label{0101f}\left\{{\begin{array}{ll}
\rho_\pm (\partial_t v_\pm+v_\pm\cdot\nabla v_\pm)+\mm{div}\mathcal{S}  =-g\rho_\pm e_3&\mbox{ in } \Omega_\pm(t),\\[1mm]
\mm{div}v_\pm=0 &\mbox{ on }\Omega_\pm(t) ,\\
d_t+v_1 \partial_1d+v_2 \partial_2d=v_3 &\mbox{ on }\Sigma(t),\\
   \llbracket v_\pm  \rrbracket =0,\
   \llbracket   \mathcal{S}_\pm \nu     \rrbracket =\vartheta \mathcal{C} \nu   &\mbox{ on }\Sigma(t) ,\\
   v_\pm=0 & \mbox{ on }\Sigma_\pm,\\
     v_\pm|_{t=0}=v^0_\pm  & \mbox{ in } \Omega_\pm (0),\\
      d|_{t=0}=d^0 & \mbox{ on } \Sigma(0).
     \end{array}}  \right.
\end{equation}

The momentum equations in \eqref{0101f}$_1$ describe the motion of the both upper heavier and lower lighter viscous fluids driven by the gravitational field along the negative $x_3$-direction, which occupy the two time-dependent disjoint open subsets $\Omega_+(t)$ and $\Omega_-(t)$ at time $t$, respectively.
Moreover the fluids are incompressible due to  \eqref{0101f}$_2$.  The two fluids interact with each other by the motion equation of a free interface \eqref{0101f}$_3$ and the interfacial jump conditions in \eqref{0101f}$_4$. The first jump condition in \eqref{0101f}$_4$ represents that the velocity is continuous across the interface.
The second jump in \eqref{0101f}$_4$ represents that the jump in the normal stress
is proportional to the mean curvature of the surface multiplied by the normal to the surface. The non-slip boundary condition of the velocities on the both upper and lower fixed flat boundaries are described by \eqref{0101f}$_5$.  \eqref{0101f}$_6$ and \eqref{0101f}$_7$ represents the initial status of the two fluids. Next we shall further explain the notations in \eqref{0101f} in details.

The subscripts $+$ resp. $-$ in the notations $f_+$ resp. $f_-$ mean that  functions, parameters or domains $f_+$ resp. $f_-$  are relevant to the upper resp. lower fluids.
For each given $t>0$, $d:=d(x_{\mm{h}},t):\mathbb{T}\mapsto (-h_-,h_+)$ is a height function of a point at the interface of stratified fluids, where $h_-$, $h_+>0$, $\mathbb{T}:=\mathbb{T}_1\times \mathbb{T}_2$, $ \mathbb{T}_i=2\pi L_i(\mathbb{R}/\mathbb{Z})$, and $2\pi L_i$ ($i=1$, $2$) are the periodicity lengths.
  The domains $\Omega_\pm(t)$ and the interface $\Sigma(t)$ are defined as follows:
$$
\begin{aligned}
&\Omega_+(t):=\{(x_{\mm{h}},x_3)~|~x_{\mm{h}}:=(x_1,x_2)\in \mathbb{T}  ,\ d(x_{\mm{h}},t)< x_3<h_+\},\\
& \Omega_-(t):=\{(x_{\mm{h}},x_3)~|~x_{\mm{h}}\in \mathbb{T} ,\ -h_-< x_3<d(x_{\mm{h}},t)\},\\
& \Sigma(t):=\{(x_{\mm{h}},x_3)~|~x_{\mm{h}}\in \mathbb{T},\ x_3:=d(x_{\mm{h}},t)\}.
\end{aligned}$$
In addition, $\Sigma_+=\mathbb{T}\times \{h_+\}$, $\Sigma_-=\mathbb{T}\times \{-h_-\}$ and we call $\Omega:=\mathbb{T}\times (-h_-,h_+)$ the domain of stratified fluids.

For given $t>0$,  $v_\pm(x,t): \Omega_\pm(t)\mapsto \mathbb{R}^3$ are the   velocities of the two fluids, and  $\mathcal{S}_\pm   $  the
stress tensors  enjoying the following expression:
\begin{equation}
\label{201901161456}
\mathcal{S}_\pm:=p_\pm I-\mu_\pm\mathbb{D}v \mbox{ with }\mathbb{D}v= \nabla v_\pm+\nabla v_\pm^{\mm{T}} .\end{equation}
In the above expression the superscript $\mm{T}$ means matrix transposition and $I$ is the $3\times 3$ identity matrix. $\rho_\pm$ are the density constants, and  the constants $\mu_\pm>0$ the shear viscosity coefficients. $g$ and $\vartheta$ represent
 the gravitational constant and  the surface tension coefficient, reps. In addition, $e_3:=(0,0,1)^{\mm{T}}$.

For a function $f$ defined on $\Omega(t)$, we define $\llbracket  f_\pm  \rrbracket:=f_+|_{\Sigma(t)}
-f_-|_{\Sigma(t)}$, where $f_\pm|_{\Sigma(t)}$ are the traces of the quantities $f_\pm$ on $\Sigma(t)$.
 $\nu$ is the unit outer normal vector at boundary $\Sigma(t)$ of $\Omega_-(t)$, and $\mathcal{C}$ the twice of the mean curvature of the internal  surface $\Sigma(t)$, i.e.,
$$
\mathcal{C}:=\frac{\Delta_{\mm{h}}d+(\partial_1 d)^2\partial_2^2
d+(\partial_2d)^2\partial_1^2d-2\partial_1d\partial_2d\partial_1\partial_2d}
{(1+(\partial_1d)^2+(\partial_2d)^2)^{3/2}}.
$$

Now we further introduce the indicator function $\chi_{\Omega_\pm(t)}$ and denote
$$
\begin{aligned}
&\rho=\rho_+\chi_{\Omega_+(t)} +\rho_-\chi_{\Omega_-(t)},\
 \mu =\mu_+ \chi_{\Omega_+(t)} +\mu_- \chi_{\Omega_-(t)},  \\
& v=v_+\chi_{\Omega_+(t)} +v_-\chi_{\Omega_-(t)},\
p=p_+\chi_{\Omega_+(t)} +p_-\chi_{\Omega_-(t)}, \\
& v^0=v_+^0\chi_{\Omega_+(0)} +v_-^0\chi_{\Omega_-(0)} ,\
 \mathcal{S}   :=p I-\mu \mathbb{D} v.
\end{aligned}$$
then the model \eqref{0101f}  can be rewritten  as  follows:
\begin{equation}\label{0103nxxxxx}\left\{{\begin{array}{ll}
\rho ( v_t+v\cdot\nabla v)+\mm{div}\mathcal{S} =-g\rho e_3 &\mbox{ in }  \Omega(t),\\[1mm]
\mm{div} v=0&\mbox{ in } \Omega(t),\\[1mm]
d_t+v_1 \partial_1d+v_2 \partial_2d=v_3 &\mbox{ on }\Sigma(t),\\
   \llbracket v  \rrbracket=0,\  \llbracket \mathcal{S} \nu\rrbracket =\vartheta \mathcal{C} \nu  \ &\mbox{ on }\Sigma(t) ,  \\
v =0 &\mbox{ on }\Sigma_-^+,\\
 v |_{t=0}= v^0  &\mbox{ in } \Omega(0),\\
d|_{t=0}=d^0  &\mbox{ on } \Sigma(0),
\end{array}}\right.\end{equation}
where we have defined that $\Omega(t):=\Omega_+(t)\cup\Omega_-(t)$, $\Sigma_-^+:=\Sigma_-\cup \Sigma_+$ and omitted the subscript $\pm$ in $\llbracket  f_\pm  \rrbracket$ for simplicity.

\subsection{Reformulation in Lagrangian coordinates}

Next we  adopt the transformation of Lagrangian coordinates so that the interface and the domains stay fixed in time.

We define that
\begin{align}
&\Omega_+:=\{(y_{\mm{h}},y_3)\in \mathbb{R}^3~|~y_{\mm{h}} \in \mathbb{T},\;\; 0<y_3<h_+\},\nonumber \\  &\Omega_-:=\{(y_{\mm{h}},y_3)\in \mathbb{R}^3~|~y_{\mm{h}} \in \mathbb{T},\;\; -h_-<y_3<0\}, \nonumber
\end{align}
and  assume that there exist invertible mappings
\begin{equation*}\label{0113}
\zeta_\pm^0:\Omega_\pm\rightarrow \Omega_\pm(0),
\end{equation*}
such that
\begin{equation}
\label{05261240}
\Sigma(0)=\zeta_\pm^0(\Sigma),\
\Sigma_\pm=\zeta_\pm^0(\Sigma_\pm) \mbox{ and } \det \nabla\zeta_\pm^0= 1.
\end{equation}
We further define $\zeta^0:=\zeta_+^0\chi_{\Omega_+}+\zeta_-^0\chi_{\Omega_-}$, and the flow map $\zeta$ as the solution to
\begin{equation}
\left\{
            \begin{array}{ll}
\partial_t \zeta(y,t)=v(\zeta(y,t),t)&\mbox{ in }\Omega_-^+
\\
\zeta(y,0)=\zeta^0(y)&\mbox{ in }\Omega_-^+,
                  \end{array}    \right.
\label{201811191443}
\end{equation}
where $\Omega_-^+:=\Omega_+\cup \Omega_-$.
We denote the Eulerian coordinates by $(x,t)$ with $x=\zeta(y,t)$,
whereas the fixed $(y,t)\in \Omega_-^+\times \mathbb{R}^+$ stand for the
Lagrangian coordinates.

In order to switch back and forth from Lagrangian to Eulerian coordinates, we shall assume that
$\zeta_\pm(\cdot ,t)$ are invertible and $\Omega_{\pm}(t)=\zeta_{\pm}(\Omega_{\pm},t)$, and since $v_\pm$ and $\zeta_\pm^0$ are all continuous across $\Sigma$, we have $\Sigma(t)=\zeta_\pm(\Sigma,t)$. In view of the non-slip boundary condition $v|_{\Sigma_-^+}=0$, we have
$$y=\zeta(y, t)\mbox{ on }\Sigma_-^+.$$

Now we set the Lagrangian unknowns
\begin{equation*}
(u,\sigma)(y,t)=(v,p+gx_3)(\zeta(y,t),t)\;\;\mbox{ for } (y,t)\in \Omega_-^+ \times(0,\infty),
\end{equation*}
then the problem \eqref{0103nxxxxx} can be rewritten as an initial-boundary value problem with an interface for $(\zeta,u)$  in Lagrangian coordinates:
\begin{equation}\label{201611041430M}\left\{\begin{array}{ll}
\zeta_t=u &\mbox{ in } \Omega_-^+,\\[1mm]
\rho u_t+\nabla_{\mathcal{A}}\sigma-\mu \Delta_{\mathcal{A}}u    =0&\mbox{ in }  \Omega_-^+,\\[1mm]
\mm{div}_{\mathcal{A}}u=0 &\mbox{ in } \Omega_-^+, \\[1mm]
 \llbracket \zeta  \rrbracket= \llbracket u  \rrbracket=0 ,\
\llbracket ((\sigma-g\zeta_3) I -\mathbb{D}_{\mathcal{A}}u)  \vec{n}  \rrbracket= \vartheta \mathcal{H}\vec{n}  &\mbox{ on }\Sigma,\\
  (\zeta,u)=(y,0) &\mbox{ on }\Sigma_-^+,\\
(\zeta,  u)|_{t=0}=(\eta^0,  u^0),  &\mbox{ in }  \Omega_-^+,
\end{array}\right.\end{equation}
where we have defined that
\begin{align}
&  \Sigma:=\mathbb{T} \times \{0\},\  \vec{n}:= {\mathcal{A}e_3}/{|\mathcal{A}e_3|},\
   \mathbb{D}_{\mathcal{A}}  u= \nabla_{\mathcal{A}} u + \nabla_{\mathcal{A}} u^\mm{T} ,\nonumber \\
&\mathcal{H}:=|\partial_1 \zeta|^2\partial_2^2\zeta-2(\partial_1\zeta\cdot \partial_2\zeta)\partial_1\partial_2\zeta+
|\partial_2\zeta|^2\partial_1^2\zeta\cdot \vec{n}/(|\partial_1\zeta|^2|\partial_2\zeta|^2-|\partial_1\zeta\cdot \partial_2\zeta|^2).\nonumber
\end{align}

We shall introduce the notations involving $\mathcal{A}$. The matrix $\mathcal{A}:=(\ml{A}_{ij})_{3\times 3}$ is defined via
\begin{equation*}
\ml{A}^{\mm{T}}=(\nabla \zeta)^{-1}:=(\partial_j \zeta_i)^{-1}_{3\times 3},
\end{equation*}  where  $\partial_{j}$ denote the partial derivative with respect to the $j$-th components of variables $y$. $\tilde{\mathcal{A}}:= \mathcal{A}-I$, and $I$ is the $3\times 3$ identity matrix.
The differential operator $\nabla_{\ml{A}}$ is defined by $$\nabla_{\ml{A}}w:=(\nabla_{\ml{A}}w_1,\nabla_{\ml{A}}w_2,\nabla_{\ml{A}}w_3)^{\mm{T}}
\;\mbox{ and }\;\nabla_{\ml{A}}w_i:=(\ml{A}_{1k}\partial_kw_i,
\ml{A}_{2k}\partial_kw_i,\ml{A}_{3k}\partial_kw_i)^{\mm{T}}$$
for vector function $w:=(w_1,w_2,w_3)$, and the differential operator $\mm{div}_\ml{A}$ is defined by
\begin{equation*}
\mm{div}_{\ml{A}}(f^1,f^2,f^3)=(\mm{div}_{\ml{A}}f^1,\mm{div}_{\ml{A}}f^2,
\mm{div}_{\ml{A}}f^3)^{\mm{T}}
\mbox{ and }\mm{div}_{\ml{A}}f^i:=\ml{A}_{lk}\partial_k f_{l}^i
\end{equation*}
for vector function $f^i:=(f_{1}^i,f_{2}^i,f_{3}^i)^{\mm{T}}$. It should be noted that we have used the Einstein convention of summation over repeated indices. In addition, we define   $\Delta_{\mathcal{A}}X:=\mm{div}_{\ml{A}}\nabla_{\ml{A}}X$.

\subsection{Linearized motion}

 We choose a constant $\bar{d}\in (-l,\tau)$. Without loss of generality, we assume that $\bar{d}=0$.  Then we consider an RT equilibrium state
 \begin{equation}
\label{201611051547}
\left\{\begin{array}{ll}
\nabla \bar{p}_\pm =-\bar{\rho}_\pm g e_3 &\mbox{ in } \Omega_\pm,\\[1mm]
     \llbracket  \bar{p}_\pm  \rrbracket e_3=0  &\mbox{ on }\Sigma,
  \end{array}\right.
\end{equation}
where $\rho$ satisfies the RT (jump) condition
 \begin{equation}
\label{201612291257}\llbracket {\rho} \rrbracket>0\mbox{ on }\Sigma.\end{equation}
  Let $\bar{p}:=\bar{p}_+\chi_{\Omega_+} +\bar{p}_-\chi_{\Omega_-}$. Then  $(v,p)=(0,\bar{p})$ with $d=0$ is  an RT equilibria solution of \eqref{0103nxxxxx}.

Denoting the perturbation in Lagrangian coordinates
 $$\eta:=\zeta-y, \ u=u-0\mbox{ and }q=\sigma-(\bar{p}(\zeta_3)+g\rho\zeta_3),$$
then subtracting \eqref{201611051547} from \eqref{201611041430M} yields the perturbation RT problem in Lagrangian coordinates:
\begin{equation}\label{201811242002xxxx}
\left\{\begin{array}{ll}
 \eta_t=  u &\mbox{ in }\Omega_-^+, \\[1mm]
 {\rho} u_{t}+ \nabla q-\mu\Delta u =\mathcal{N}_1& \mbox{ in }\Omega_-^+,\\[1mm]
 \mm{div}u=\mathcal{N}_2& \mbox{ in }\Omega_-^+,\\[1mm]
\llbracket \eta  \rrbracket= \llbracket u  \rrbracket=0,\
  \llbracket((q-g\rho \eta_3)I-\mu \mathbb{D}u) e_3  \rrbracket =\vartheta\Delta_{\mm{h}}\eta_3e_3+\mathcal{N}_3  &\mbox{ on }\Sigma,\\
 (\eta,u)= 0 &\mbox{ on }\Sigma_-^+,\\
(\eta, u)|_{t=0}= (\eta^0, u^0)  &\mbox{ in }  \Omega_-^+,\end{array}\right.\end{equation}
where $\Delta_{\mm{h}}:=\partial_1^2+\partial_2^2$ and the nonlinear terms $\mathcal{N}_1$--$\mathcal{N}_3$  are defined as follows:
\begin{align}
& \mathcal{N}_1= \mm{div}_{\tilde{\ml{A}}}\nabla_{\mathcal{A}} u+\mm{div}\nabla_{\tilde{\ml{A}}} u-\nabla_{\tilde{\mathcal{A}}}q,\ \mathcal{N}_2=-  \mm{div}_{ \tilde{\mathcal{A}}} u,\nonumber \\
&\mathcal{N}_3= \mathbb{D}_{\tilde{\ml{A}}} u+\vartheta \mathcal{H}\vec{n} -\vartheta\Delta_{\mm{h}}\eta_3e_3 .\nonumber
\end{align}
Omitting the nonlinear terms in \eqref{201811242002xxxx}, we get a linearized RT problem:
\begin{equation}\label{201811242002xx}
\left\{\begin{array}{ll}
 \eta_t=  u &\mbox{ in }\Omega_-^+, \\[1mm]
\rho u_{t}+ \nabla q-\mu\Delta u  =0 & \mbox{ in }\Omega_-^+,\\[1mm]
\div u  =0 &\mbox{ in }\Omega_-^+,\\
\llbracket \eta  \rrbracket= \llbracket u  \rrbracket=0&\mbox{ on }\Sigma,\\
\llbracket ((q-g\rho \eta_3)I-\mu \mathbb{D}u) e_3 \rrbracket =\vartheta\Delta_{\mm{h}}\eta_3e_3 &\mbox{ on }\Sigma,\\
 (\eta,u)= 0 &\mbox{ on }\Sigma_-^+,\\
(\eta,u)|_{t=0}= (\eta^0,u^0)  &\mbox{ in } \Omega_-^+.\end{array}\right.\end{equation}
Of course, the motion equations of stratified viscous fluids in linear stage can be approximatively described by \eqref{201811242002xx}.

The inhibition of RT instability by surface tension was first analyzed by Bellman--Phennington \cite{BRPRHEQAM1954} based on a linearized two-dimensional (2D) motion equations of stratified incompressible inviscid fluids defined on the domain $2\pi L_1\mathbb{T}_1\times (-h_-,h_+)$ (i.e., $\mu=0$ in the corresponding 2D case of \eqref{201811242002xx}) in 1953.
More precisely, they proved that the linear 2D  stratified incompressible inviscid fluids is stable, resp. unstable for $\vartheta>g\llbracket\rho\rrbracket L_1^2$, resp. $\vartheta<g\llbracket\rho\rrbracket L_1^2$.  The value $g\llbracket\rho\rrbracket L_1^2$ is a threshold of surface tension coefficient for linear stability and linear instability.
Similar result was also found in the 3D viscous case, for example, Guo--Tice proved that $\vartheta_{\mm{c}}:={ g\llbracket\rho\rrbracket\max\{L_1^2,L_2^2\} }$ is a threshold of surface tension coefficient for stability and instability in the linearized 3D stratified  compressible viscous fluids defined on $\Omega$ \cite{GYTI2}. Next we further review the mathematical progress for the nonlinear case.

Pr$\mathrm{\ddot{u}}$ess--Simonett first  proved that the RT equilibria solution
of the stratified incompressible viscid fluids defined on the domain $\mathbb{R}^3$ is unstable
based on a Henry instability method \cite{PJSGOI5x}.
Later, Wang--Tice--Kim verified that the RT equilibria solution
of stratified incompressible viscous fluids defined on $\Omega$ is stable, resp. unstable for $\vartheta >\vartheta_{\mathbb{T}} $, resp. $ \vartheta \in [0, \vartheta_{\mathbb{T}})$  \cite{WYJTIKCT,wang2011viscous}. Jang--Wang--Tice further obtained the same results of stability and instability in the corresponding compressible case \cite{JJTIWYHCMP,JJTIWYHCMP121}. Recently, Wilke also proved there exists a threshold $\vartheta_{\mm{c}} $ for the stability and instability of stratified viscous fluids (with heavier fluid over lighter fluid) defined on a cylindrical domain with finite height \cite{WMRTITNS2017}. Finally, we mention that the results of nonlinear RT instability in inhomogeneous fluid (without interface) were obtained based on the classical bootstrap instability method, see \cite{HHJGY}, resp. \cite{JFJSO2014} for inviscid, resp. viscous cases.

\section{Main result}\label{201806101821}

In this paper, we investigate the effect of surface tension on the linear RT instability by the linearized motion \eqref{201811242002xx}. Wang--Tice used  discrete
Fourier transformation  and modified variational method of ODEs to prove the existence of growth solutions with a largest growth rate $\Lambda_\vartheta$ for \eqref{201811242002xx} with $h_+=1$ under the condition
$\vartheta_\vartheta \in (0,\vartheta_{\mathbb{T}})$  \cite{wang2011viscous}.
Moreover, they provided an upper bound for $\Lambda_\vartheta$:
$$\Lambda_\vartheta \leqslant {h_-g \llbracket \rho  \rrbracket}/{4\mu_-}\mbox{ for }h_+=1.$$

In this paper, we exploit  modified variational method of PDEs and existence theory of stratified (steady) Stokes problem to prove  the existence of growth solutions with a largest growth rate $\Lambda_\vartheta$ for \eqref{201811242002xx} under the instability condition   $\vartheta \in [0,\vartheta_{\mathbb{T}})$.
Moreover we  find a new upper bound:
\begin{equation}
\label{201901251413}
\Lambda_{\vartheta}\leqslant m:=\min\left\{\frac{  (\vartheta_{\mm{c}}-\vartheta) }{4\max\{L_1^2,L_2^2\}}
\min \left\{\frac{h_+}{\mu_+},\frac{{h}_-}{\mu_-}\right\},\left(\frac{4(g\llbracket\rho\rrbracket (\vartheta_{\mm{c}}-\vartheta))^2 }{\vartheta_{\mm{c}}^2 \max\{\rho_+\mu_+ ,
\rho_-\mu_-\}}\right)^{\frac{1}{3}}\right\}.
\end{equation}

It is easy to see that
$$m  \leqslant {h_-g \llbracket \rho  \rrbracket}/{4\mu_-}. $$
Therefore our upper bound is more precise than Wang--Tice's one.
Moreover, we see from \eqref{201901251413} that
\begin{equation}
\label{201901272330}
\Lambda_\vartheta\to 0\mbox{ as }\vartheta\to \vartheta_{\mm{c}}.
\end{equation}
In classical Rayleigh--Taylor (RT) experiments \cite{GJSAMW,HSWWNH}, it has been shown the phenomenon of that the instability growth is limited by surface tension during the linear stage, where the growth is exponential in time. Obviously, the convergence behavior \eqref{201901272330} mathematically verifies the phenomenon.

Before stating our main results in detail, we shall introduce some simplified notations throughout this article.
\begin{enumerate}
  \item Basic notations:
$I_T:=(0,T)$.  $\mathbb{R}_+:=(0,\infty)$. The $j$-th difference quotient of size $h$ is $D_j^h w:=(w(y+h e_j)-w(y))/h$ for $j=1$ and $2$, and $D^h_{\mm{h}}w:=(D_1^hw_1,D_2^hw_2)$, where $|h|\in (0,1)$.
 $\Re f $, reps. $\Im f$ denote the real, resp. imaginary parts of the
complex function $f$.
 $\nabla_{\mm{h}}^k f$ denotes a $(k+1)\times (k+1)$ matrix   $(\partial_1^{i}\partial_2^j f)_{ij}$ for $ k\geqslant0 $.
 $a\lesssim b$ means that $a\leqslant cb$ for some constant $c>0$, where the positive constant $c$ may depend on the domain $\Omega$, and known parameters such as $\rho_\pm$, $\mu_\pm$, $g$ and $\vartheta$, and may vary from line to line. 
  \item Simplified  norms:
$ \|\cdot \|_{i} :=\|\cdot \|_{W^{i,2}}$, $|\cdot|_{s} := \|\cdot|_{\Sigma} \|_{H^{s}(\mathbb{T})}$,
where  $s$ is a real number, and $i$   a non-negative integer.
\item Functionals: $
  \mathcal{E}(w):= \vartheta|\nabla_{\mm{h}}w_3|_{0}^2-g\llbracket\rho\rrbracket |w_3|_{0}^2$ and $
 \mathcal{F}(w,s):=  -(\mathcal{E}(w) +s\|\sqrt{\mu} \mathbb{D}w\|^2_0/2)$.
\end{enumerate}

In addition, we shall give the definition of the largest growth rate of RT instability in the linearized  RT problem.
\begin{definition}
\label{201804072001}
We call $\Lambda>0$ the largest growth rate of RT instability in the linearized  RT problem \eqref{201811242002xx},
if it satisfies the following two conditions:
\begin{enumerate}
  \item[(1)] For any strong solution $(\eta, u)\in C^0([0,T),H^3\cap H^2_\sigma)\cap L^2(I_T,H^3\cap H^3_\sigma)$ of the linearized  RT problem with  $q$ enjoying  the regularity $ q\in  C^0([0,T),H^1)\cap L^2(I_T,H^2)$,
      then we have, for any $t\in [ 0,T)$,
\begin{align}
&\label{estfsharprate}
\| (\eta,u) \|_{1}^2+\| u_t \|^2_{0 }+ \int_0^t\| u(s) \|^2_{1}\mm{d}s\lesssim
  e^{2\Lambda t}(\|\eta^0\|_3^2+\|u^0\|_2^2) .
\end{align}
  \item[(2)] There exists a strong solution $(\eta, u)$ of the linearized RT problem in the form
  $$
(\eta, u):=e^{\Lambda t}( {u}, {\eta} ),$$
where $( \eta, {u})\in H^2$.
\end{enumerate}
\end{definition}

Now we state the first result on the existence of largest growth rate in the linearized RT problem.
\begin{thm}
\label{201806012301xxxxxxxx}
Let  $g>0$, $\rho>0$ and $\mu>0$ are given. Then, for any given \begin{equation}
\label{201811241424}
 \vartheta \in [0, \vartheta_{\mm{c}}:=g\llbracket \rho \rrbracket{\max\{L_1^2,L_2^2\}}).
\end{equation}
there is an unstable solution
$$( \eta, u,q):=e^{\Lambda t}( w/\Lambda,w,\beta)$$
 to  the linearized RT problem \eqref{201811242002xx}, where $(w,\beta)\in   H^\infty$
 solves  the boundary value problem:
\begin{equation}
\label{201604061413}
\left\{\begin{array}{ll}
\Lambda^2\rho w=    \Lambda(\mu \Delta w-\nabla\beta) &\mbox{ in }\Omega_-^+, \\ \mm{div}w=0 &  \mbox{ in }\Omega_-^+, \\
   \llbracket w \rrbracket= 0,\
   \llbracket  \Lambda (\beta I-\mu \mathbb{D}w)e_3
        - g \rho w_3 e_3 \rrbracket =  \vartheta\Delta_{\mm{h}}w_3  e_3  &\mbox{ on }\Sigma, \\
 w = {0}& \mbox{ on }\Sigma_-^+  \end{array} \right.
\end{equation}
with a largest growth rate  $\Lambda>0$ satisfying
\begin{equation}
\Lambda^2= \sup_{\varpi\in \mathcal{A} }F(\varpi, \Lambda)=F(w, \Lambda) .
\label{201901251824} \end{equation}
Moreover,
\begin{align}
&
w_3\neq 0,\ \partial_3 w_3\neq 0,\  \mm{div}_{\mm{h}}w_{\mm{h}}\neq 0\mbox{ in }\Omega_-^+, \ |w_3|\neq 0\mbox{ on }\Sigma.
\label{201602081445MH}
\end{align}
\end{thm}

Next we briefly introduce how to  prove Theorem \ref{201806012301xxxxxxxx} by modified variational method of PDEs and regularity theory of stratified (steady) Stokes problem. The detailed proof will be given in Section \ref{201805302028}.

We  assume a growing mode ansatz to the linearized problem:
\begin{equation*}
\eta(x,t)=\tilde{\eta}( x)e^{\Lambda t},\  u(x,t)=w( x)e^{\Lambda  t},\
q( x,t)=\beta( x)e^{\Lambda  t}
\end{equation*}  for some $\Lambda >0$.
Substituting this ansatz into the linearized RT problem  \eqref{201811242002xx}, we get a spectrum problem
\begin{equation*}
\left\{\begin{array}{ll}
\Lambda \tilde{\eta}=w &\mbox{ in }\Omega_-^+, \\
\Lambda \rho w=  \mu \Delta w-\nabla\beta &\mbox{ in }\Omega_-^+, \\
\mm{div}w=0 &  \mbox{ in }\Omega_-^+, \\
   \llbracket w \rrbracket= 0,\
   \llbracket (\beta I-\mu \mathbb{D}w)e_3
        - g \rho \tilde{\eta}_3 e_3 \rrbracket =  \vartheta\Delta_{\mm{h}}\tilde{\eta}_3  e_3  &\mbox{ on }\Sigma, \\
 w = {0}& \mbox{ on }\Sigma_-^+  \end{array} \right.
   \end{equation*}
and then eliminating $\tilde{\eta}$  by using the first equation, we arrive at
the boundary-value problem \eqref{201604061413}
for $w$ and $\beta$. Obviously, the linearized RT problem
is unstable, if there exists a solution $(w,\beta)$ to the boundary-value problem \eqref{201604061413} with $\Lambda>0$.

To look for the solution,  we use a modified variational method of PDEs, and thus modify \eqref{201604061413} as follows:
 \begin{equation}\label{0301nn}
 \left\{\begin{array}{ll}
\alpha\rho w=   s(\mu \Delta w-\nabla\beta) &\mbox{ in }\Omega_-^+, \\ \mm{div}w=0 &  \mbox{ in }\Omega_-^+, \\
   \llbracket w \rrbracket= 0,\
   \llbracket  s (q I-\mu \mathbb{D}w)e_3
        - g \rho w_3 e_3 \rrbracket =  \vartheta\Delta_{\mm{h}}w_3  e_3  &\mbox{ on }\Sigma, \\
 w = {0}& \mbox{ on }\Sigma_-^+  \end{array} \right.   \end{equation}
where $s>0$ is a parameter. To emphasize the  dependence of $s$ upon $\alpha$ and $\vartheta$, we will   write
$ \alpha(s,\vartheta)=\alpha$.

Noting that the modified problem \eqref{0301nn} enjoys
 the following variational identity
 $$\alpha(s,\vartheta)\|\sqrt{ {\rho}}w\|_0^2= \mathcal{F}(w,s).$$
Thus, by a standard  {variational approach}, there exists a maximizer $w\in \mathcal{A} $ of  the functional $\mathcal{F}$ defined on $\mathcal{A}$; moreover $w$ is  just a weak solution to \eqref{0301nn} with $\alpha$   defined by the relation
\begin{equation}\label{0206nn}
 \alpha( s  ,\vartheta) =\sup_{w\in\mathcal{A} }\mathcal{F}(w,s )\in \mathbb{R},\end{equation}
 see Proposition \ref{201811241541}.
Then we further use the method of difference quotients and the existence theory of the stratified (steady) Stokes problem
to improve the regularity of the weak solution, and thus prove that $(w,\beta)\in H^\infty$ is a classical solution to the boundary-value problem \eqref{0301nn}, see Proposition \ref{201901281636}.

In view of the definition of $\alpha(s ,\vartheta)$ and the instability condition \eqref{201811241424}, we can infer that, for given $\vartheta$, the function $\alpha (s,\cdot)$ on the variable $s$ enjoys some good properties (see Proposition \ref{201901261856}), which imply that  there exists
  a $\Lambda $ satisfying the fixed-point relation
\begin{equation} \label{growthn}
 \Lambda =\sqrt{\alpha(\Lambda ,\cdot)}  \in (0,\mathfrak{S}_\vartheta). \end{equation}
Then we obtain a nontrivial solution $(w,\beta)\in H^\infty$ to \eqref{201604061413}
  with $\Lambda $ defined by \eqref{growthn}, and therefore the linear instability follows. Moreover,
  $\Lambda $ is  the largest growth rate of RT instability in the linearized  RT problem (see Proposition \ref{bestgrowth}), and thus we get  Theorem \ref{201806012301xxxxxxxx}.

Next we turn to introduce the second result on the properties of largest growth rate constructed by
\eqref{growthn}.
\begin{thm}
\label{201901281046}
The largest growth rate $\Lambda_{\vartheta}:=\Lambda$ in Theorem \ref{201806012301xxxxxxxx}  enjoys the estimate \eqref{201901251413}. Moreover,
\begin{equation}\label{201901252005}
\Lambda_\vartheta\mbox{ strictly decreases and is continuous with respect to }\vartheta\in [0,\vartheta_{\mm{c}}).
\end{equation}
In particular, we have $\Lambda_{\vartheta}\rightarrow 0$ as $\vartheta\rightarrow \vartheta_{\mm{c}}$.
\end{thm}

The proof of Theorem \ref{201901281046} will be presented in  Section \ref{201901292122}.
Here we briefly mention the  idea of proof.
We find that, for fixed $s$,
  $\alpha(\cdot,\vartheta)$ defined by \eqref{0206nn} strictly  decreases  and is continuous with respect to
  $\vartheta$ (see Proposition \ref{lem:0201}). Thus, by the fixed-point relation \eqref{growthn} and some analysis  based on the definition of continuity, we can show that $\Lambda_\vartheta:=\Lambda$ also inherits the monotonicity and continuity of $\alpha(\cdot,\vartheta)$. Finally, we derive \eqref{201901251413} from \eqref{201901251824} by some estimate techniques.

\section{Preliminary}

This section is devoted to the introduce of some preliminary lemmas, which will be used in the next two sections.
\begin{lem}\label{xfsddfsf2018050813379safdadf} Difference quotients and weak derivatives: Let
$D$ be $\Omega$, or $\mathbb{T}$.
\begin{enumerate}
 \item[(1)] Suppose  $1\leqslant p<\infty$ and $w\in W^{1,p}(D)$. Then
$\|D^{h}_{\mm{h}} w\|_{L^p(D) }\lesssim  \|\nabla_{\mm{h}}w\|_{L^p(D)}$.
 \item[(2)] Assume $1< p<\infty$,  $w\in L^p(D)$, and there exists a constant $c$ such that
$\|D^h_{\mm{h}} w\|_{L^p(D)}\leqslant c$. Then $\nabla_{\mm{h}}
w\in L^{p}(D)$ satisfies $\|\nabla_{\mm{h}}w\|_{L^p(D)}\leqslant c$ and
 $D^{-h_k}_{\mm{h}} w \rightharpoonup \nabla_{\mm{h}}w$ in $L^p(D)$ for some subsequence
$-h_k\to 0$.
\end{enumerate}
\end{lem}
\begin{pf}
Following the argument of  \cite[Theorem 3]{ELGP}, and use the periodicity of
$w$, we can easily get the desired conclusions. \hfill $\Box$
\end{pf}
 \begin{lem}
\label{xfsddfsf201805072212}
Existence  theory of a stratified (steady) Stokes problem (see \cite[Theorem 3.1]{WYJTIKCT}): let $k\geqslant 0$, $f^{\mm{S},1}\in H^{k}$ and  $f^{\mm{S},2}\in H^{k+1/2}$, then there exists a unique solution $(u,q)\in H^{k+2}\times \underline{H}^{k+1}$ satisfying
\begin{align}
\left\{\begin{array}{ll}
\nabla q-\mu\Delta u =f^{\mm{S},1}&\mbox{ in }  \Omega, \\[1mm]
\llbracket u \rrbracket=0,\ \llbracket (q I-\mathbb{D}u )e_3 \rrbracket =f^{\mm{S},2} &\mbox{ on }\Sigma, \\
u=0 &\mbox{ on }\Sigma_-^+.
\end{array}\right.
\label{201808311541}
\end{align}
Moreover,
\begin{equation}
\label{2011805302036}
 \| u  \|_{\mm{S},k}   \lesssim \|f^{\mm{S},1}\|_{k} +|f^{\mm{S},2}|_{k+1/2} .
\end{equation}
\end{lem}
\begin{lem}
\label{201811241500}
Equivalent form of instability condition:
the instability condition \eqref{201811241424} is equivalent to
the following integral version of instability condition:
\begin{equation}
g\llbracket\rho\rrbracket |w_3|_{0}^2-\vartheta|\nabla_{\mm{h}}w_3|_{0}^2>0 \mbox{ for some }
w\in H^1_{\sigma,3}.
\label{201811241957}
\end{equation}
\end{lem}
\begin{pf}
The conclusion in Lemma \ref{201811241500} is obvious, if we have the assertion:
\begin{equation}
\label{2018060102215680}
a:=\sup_{w\in H_{\sigma,3}^1 }
\frac{|w_3|_0^2}{
|\nabla_{\mm{h}}w_3 |_0^2}=  {\max\{L_1^2,L_2^2\} }\mbox{ for }\vartheta\neq 0.
\end{equation}
  Next we verify \eqref{2018060102215680} by two steps. Without loss of generality, we assume that $L_1^2={\max\{L_1^2,L_2^2\}}$.

(1) We first prove that  $a\geqslant L_1^2$. We choose a non-zero function $\psi\in H_0^2(-h_-,h_+)$ such that $\psi(0)\neq 0$. We denote
 $$ {w}=(\psi'  (y_3)\cos( L^{-1}_1y_1 ),0,  L^{-1}_1\psi (y_3) \sin (L^{-1}_1 y_1 )),$$ then $ {w} \in H^1_{\sigma,\vartheta}$ and
$$
\begin{aligned}
& \frac{| {w}_3|^2_0} {|\nabla_{\mm{h}}  { w}_3|^2_0}
=  \frac{   \int_0^{2\pi L_1}
 \sin^2 ( L^{-1}_1 y_1 )\mm{d}y_1  }{L^{-2}_1  \int_0^{2\pi L_1}
  \cos^2(  L^{-1}_1 y_1) \mm{d}y_1  }=L_1^2,
\end{aligned}
$$
which yields $a\geqslant L_1^2$.

(2) We turn to the proof of $a\leqslant L_1^2$.  It should be noted that
\begin{equation}\label{201806101508}
|\nabla_{\mm{h}}w_3 |_{0}^2=0\mbox{ if and only if } w_3=0\mbox{ for any given }w\in H_{\sigma,\vartheta}^1.
\end{equation}
In fact, let $w\in H_{\sigma,3}^1$.
Since $\mm{div}w=0$, we have
$$-\int_\Sigma w_3\mm{d}y_{\mm{h}}=\int_{\mathbb{T}\times (0,h_+)}\mm{div}w\mm{d}y=0.$$
Thus, using Pocare's inequality, we have
$$|w_3|_{0}\lesssim  |\nabla_{\mm{h}}w_3|_{0},$$
which immediately implies the assertion \eqref{201806101508}.

Let $w\in H_{\sigma,3}^1$.  Then
 $|\nabla_{\mm{h}} w_3 |_{0}^2\neq 0$.
 Let $\hat{w}_3(\xi,y_3)$ be the horizontal Fourier transform of $w_3(y) $, i.e.,
$$ \hat{w}_3(\xi,y_3)=\int_{\Sigma}w_3( y_{\mm{h}},y_3)e^{-\mm{i}y_{\mm{h}}\cdot\xi}\mm{d}x_{\mm{h}}, $$
where  $\xi=(\xi_1,\xi_2)$, then $\widehat{\partial_3  w_3} = \partial_{3} \widehat{w}_3$. We denote $\psi(\xi,y_3):= \psi_1(\xi,y_3) + \mm{i}\psi_2(\xi,y_3):=\hat{w}_3(\xi,y_3)$, where $\psi_1$ and $\psi_2$ are real functions.
Noting that $\psi(0)=0$, by  Parseval theorem  (see \cite[Proposition 3.1.16]{grafakos2008classical}), we have
 \begin{equation*}
\begin{aligned}
|\nabla_{\mm{h}}{w}_3 |^2_0=& \frac{1}{4\pi^2 L_1L_2}\sum_{\xi\in (L^{-1}_1\mathbb{Z}\times L^{-1}_2\mathbb{Z})}
|\xi|^2|\psi(\xi,0)|^2
\geqslant
 L^{-1}_1 |w_3|^2_0,
\end{aligned}\end{equation*}
which immediately yields that $a\leqslant L^2_1$.
 The proof is complete.
\hfill $\Box$
\end{pf}
\begin{lem}\label{xfsddfsf201805072234FRe}
Friedrichs's inequality (see \cite[Lemma 1.42]{NASII04}): Let $1\leqslant p<\infty$, and $D$ be a bounded Lipschitz domain. Let a set $\Gamma\subset \partial D$ be measurable with respect to the $(N-1)$-dimensional measure $ {\mu}:=\mm{meas}_{N-1}$ defined on $\partial D$ and let $\mm{meas}_{N-1}(\Gamma)>0$. Then 
\begin{equation*}
\|w\|_{W^{1,p}(D)}\lesssim \|\nabla w\|_{L^p(D)}^2\end{equation*}
 for all $u\in W^{1,p}(D)$ satisfying that the trace of $u$ on $\Gamma$ is equal to $0$ a.e. with respect to the $(N-1)$-dimensional measure $ {\mu}$.
\end{lem}
\begin{rem}
By Friedrichs's inequality and the fact
\begin{equation}
\label{201901272014}
\|\nabla w\|_0^2=\|\mathbb{D}w\|_0^2/2 \mbox{ for any }w\in H_\sigma^1,
\end{equation}
 we get the Korn's inequality
\begin{equation}
\label{201901291007}
\|w\|_1 \lesssim  \|\mathbb{D}w\|_0^2 \mbox{ for any }w\in H_\sigma^1.\end{equation}
\end{rem}
\begin{lem}
\label{xfsddfsf201805072254}
 Trace estimates:
\begin{align}
& |w|_{0} \leqslant  \|  w \|_1
\mbox{ for any }w\in H^{1}_\sigma,
\label{201901281645xx}\\
& |w|_{0} \leqslant  \sqrt{h_\pm/2}   \|\mathbb{D} w \|_{L^2(\Omega_\pm)}/2
\mbox{ for any }w\in H^{1}_\sigma.
\label{201901281645}
\end{align}
\end{lem}
\begin{pf}
See \cite[Lemma 9.7]{JFJSOMITNWZ} for \eqref{201901281645xx}.
Since $C_\sigma^\infty:=C^\infty_0(\mathbb{R}^2\times (-h_-,h_+))\cap H_\sigma^1$ is dense in $H^1_\sigma$, it suffices to prove that
\eqref{201901281645} holds for any $w\in C_\sigma^\infty$ by \eqref{201901281645xx}.

Let $\hat{w}$ be the horizontal Fourier transformed function of $w\in C_\sigma^\infty$,
and
$$ \varphi(\xi,y_3)=\mm{i} \hat{w}_1(\xi,y_3), \;\;
\theta(\xi,y_3)=\mm{i}\hat{w}_2(\xi,y_3),\;\;\psi(\xi,y_3)=\hat{w}_3(\xi,y_3).$$
Then
\be   \label{dive=0}
\xi_1\varphi+\xi_2\theta+\psi'=0   \ee
and $\psi (\cdot, y_3)\in H^2_0(-h_-,h_+)$, because of $\mathrm{div} {w}=0$ and $w|_{\Sigma_-^+}=0$. Moreover,
 \begin{equation*}
\widehat{\nabla  {w}}=(\widehat{\partial_i w_j})=\left(
                                                     \begin{array}{ccc}
 \xi_1\varphi &  \xi_2\varphi & -\mm{i}\varphi'\\
 \xi_1\theta&  \xi_2\theta  &-\mm{i}\theta' \\
 \mm{i}\xi_1\psi &  \mm{i}\xi_2\psi  & \psi'  \\
 \end{array}                 \right).
\end{equation*}
In addition, we can deduce from \eqref{dive=0} that
\begin{equation}
\label{201901292123}
\psi(0,y_3)=0\mbox{ for }\xi=0.
\end{equation}

 By \eqref{201901292123} and the Fubini and Parseval theorems,  one has
\begin{equation} \label{0305jk}
|w_3 |^2_0 = \frac{1}{4\pi^2 L_1L_2}\sum_{\xi\in (L^{-1}_1\mathbb{Z}\times L^{-1}_2\mathbb{Z})\backslash\{0\}} |\psi(\xi,0)|^2 \end{equation}
and
 \begin{align}
& \frac{1}{2}\| \mathbb{D}{w} \|^2_{L^2(\Omega_-)}=  \frac{1}{8\pi^2 L_1L_2}\sum_{\xi\in (L^{-1}_1\mathbb{Z}\times L^{-1}_2\mathbb{Z})} \sum_{1\leqslant i,j\leqslant 3}
\int_{-h_-}^0   |\widehat{\partial_i {w}_j}+\widehat{\partial_j {w}_i}|^2\mm{d}y_3\nonumber \\
&=\frac{1}{4\pi^2 L_1L_2}\sum_{\xi\in (L^{-1}_1\mathbb{Z}\times L^{-1}_2\mathbb{Z})\backslash\{0\}}M_1^\xi(\varphi,\theta,\psi)\nonumber
\\
&\quad +
\frac{1}{4\pi^2 L_1L_2} \int_{-h_-}^0
\left( |\varphi'(0,y_3)|^2+|\theta'(0,y_3)|^2 \right)\mm{d}y_3, \label{ueMphitheta1}
\end{align}
where
 $$\begin{aligned}
 M_1^\xi( \varphi,\theta,\psi):=&\int_{-h_-}^0
\left(|\xi|^2(|\varphi|^2+|\theta|^2+|\psi|^2)\right.\\
&\quad \left.+2\Re {\psi}'' \Re{\psi} + 2\Im{\psi}''
\Im{\psi}+|\varphi'|^2+|\theta'|^2+3|\psi'|^2\right)\mm{d}y_3.
\end{aligned}$$

Using \eqref{dive=0}, we find that
$$\begin{aligned}
& |\psi'|^2 = \xi_1^2|\varphi|^2+\xi_2^2|\theta|^2+
2 \xi_1\xi_2(\Re {\varphi}   \Re{\theta} +  \Im{\varphi} \Im{\theta}) \leqslant |\xi|^2(|\varphi|^2+ |\theta|^2), \\
&  |\psi''|^2  \leqslant |\xi|^2(|\varphi'|^2+ |\theta'|^2) ,
\end{aligned}$$
which imply that
\begin{equation}\label{0311nn}\begin{aligned}
   \int_{-h_-}^0 (4 | {\psi}'|^2+||\xi| {\psi}+ {\psi}''/|\xi||^2) \mm{d}y_3\leqslant
M_1^\xi(\varphi,\theta,\psi)
\end{aligned}\end{equation}for given $ \xi\in  (L^{-1}_1\mathbb{Z}\times L^{-1}_2\mathbb{Z})\backslash\{0\}$.
Employing \eqref{0305jk}--\eqref{0311nn} and the relation
$$
\phi^2(0)\leqslant h_-  \|\phi'\|_{L^2(-h_-,0)}^2
\mbox{ for any }\phi\in H^1_0(-h_-,h_+),$$
we obtains
       \begin{align}
 |w_3|^2_0  = &  \frac{1}{4\pi^2 L_1L_2}\sum_{\xi\in (L^{-1}_1\mathbb{Z}\times L^{-1}_2\mathbb{Z})\backslash\{0\}} |\psi(\xi,0)|^2 \nonumber  \\
\leqslant &
 \frac{h_-}{16 \pi^2 L_1L_2}
 \sum_{\xi\in (L^{-1}_1\mathbb{Z}\times L^{-1}_2\mathbb{Z})\backslash\{0\}}  \int_{-h_-}^0  (4 | {\psi}'|^2+(|\xi|  {\psi}+ {\psi}''/|\xi|)^2) \mm{d}y_3\nonumber \\
  \leqslant  &   \frac{h_-}{16  \pi^2 L_1L_2}\sum_{\xi\in (L^{-1}_1\mathbb{Z}\times L^{-1}_2\mathbb{Z})\backslash\{0\}} M_1^\xi( \varphi,\theta,\psi)
\leqslant   {h_-}  \|\mathbb{D} w\|^2_{L^2(\Omega_-)}/8.
  \label{201901271947}\end{align}
Similarly, we also have
$$ |w_3|^2_0 \leqslant   {h_+}  \|\mathbb{D} w\|^2_{L^2(\Omega_+)} /8,$$
which, together with \eqref{201901271947}, yields the desired conclusion.
This completes the proof.  \hfill$\Box$
\end{pf}
\begin{rem}
From the derivation of \eqref{201901281645}, we easily see that
\begin{equation}
\|\partial_3 w_3\|_{L^2(\Omega_\pm)}^2\leqslant \|\mathbb{D} w\|_{L^2(\Omega_\pm)}^2/8\mbox{ for any }w\in H^1_\sigma.
\label{201901301148}
\end{equation}
\end{rem}
\begin{lem}
Negative trace estimate:
 \begin{equation}\label{06201929201811041930}|u_3|_{-1/2}\lesssim \|u\|_0+\|\mm{div}u\|_0\;\;\;\mbox{ for any }u:=(u_1,u_2,u_3)\in H_0^1.
\end{equation}
\end{lem}
\begin{pf}
Estimate \eqref{06201929201811041930} can be derived by integration by parts and an inverse trace theorem  \cite[Lemma 1.47]{NASII04}. \hfill $\Box$
\end{pf}
\begin{lem}
\label{201901251652}
Let $X$ be a given Banach space with dual $X^*$ and let $u$ and $w$ be two functions belonging to $L^1((a,b),X)$. Then the following two conditions are equivalent
\begin{enumerate}
  \item[(1)] For each test function $\phi\in C_0^\infty(a,b)$,
  $$\int_a^b u(t)\phi'(t)\mm{d}t=-\int_a^b w(t)\phi(t)\mm{d}t.$$
  \item[(2)]
  For each $\eta\in X^*$,
  $$\frac{\mm{d}}{\mm{d}t}<u,\eta>_{X\times X^*}=<w,\eta>_{X\times X^*},$$
  in the scalar distribution sense, on $(a,b)$, where $<\cdot ,\cdot>_{X\times X^*}$ denotes the dual pair between $X$ and $X^*$.
\end{enumerate}
\end{lem}
\begin{pf}
See Lemma 1.1 in Chapter 3 in \cite{TRNSETNAxxx}.
\end{pf}

\section{Linear instability}\label{201805302028}

In this section, we will use modified variational method to construct unstable solutions for the linearized RT problem. The modified variational method  was firstly used by Guo and Tice  to  construct unstable solutions to a class of ordinary differential equations arising from a linearized RT instability problem \cite{GYTI2}. In this paper, we directly apply Guo and Tice's modified variational method to  the partial differential equations \eqref{201604061413}, and thus  obtain a linear instability result of the RT problem by further using an existence theory  of stratified Stokes problem. Next we prove  Theorem \ref{201806012301xxxxxxxx} by four subsections.

\subsection{Existence of weak solutions to the modified problem}

In this subsection, we consider the existence of weak solutions to the modified problem
 \begin{equation}
\label{2016040614130843xd}
\left\{    \begin{array}{ll}
\nabla q- s\mu \Delta w
 =   \alpha(s,\vartheta) {\rho} w,\ \mm{div}w=0&\mbox{ in }\Omega_-^+, \\
\llbracket w  \rrbracket=0,\
\llbracket  ((q-g\rho w_3) I- s \mu\mathbb{D}  w)e_3 \rrbracket=  \vartheta\Delta_{\mm{h}}w_3  e_3&\mbox{ on }\Sigma, \\
  w = {0}& \mbox{ on }\Sigma_-^+,  \end{array} \right.
                       \end{equation}
where $s>0$ is any given.
To prove the existence of weak solutions of the above problem, we consider the variational problem of the functional $F(\varpi,s )$:
\begin{equation}
\label{0206}
\alpha(s,\vartheta): =\sup_{\varpi\in\mathcal{A}} F(\varpi,s )
\end{equation}
for given $s>0$, where we have defined that
$$F(\varpi,s):= -(\mathcal{E}(w)+s \|\sqrt{\mu}\mathbb{D}w\|_0^2/2).$$
 Sometimes, we denote $\alpha(s,\vartheta)$ and $F(\varpi,s)$ by $\alpha$ (or $\alpha(s)$) and $F(\varpi)$ for simplicity, resp..
Then we have the following conclusions.
\begin{pro}
\label{201811241541}
Let $s>0$ be any given.
\begin{enumerate}[(1)]
\item In the variational problem \eqref{0206}, $F(\varpi)$ achieves its supremum on $\mathcal{A}$.
\item Let $w$ be a maximizer and $\alpha:= \sup_{\varpi\in\mathcal{A}}F(\varpi) $, the  $ w$ is a weak solution the boundary problem \eqref{2016040614130843xd} with given $\alpha$.
 \end{enumerate}
\end{pro}
\begin{pf}
Noting that
\begin{equation}
\label{201805072235}
|v|_0^2 \lesssim \|v\|_0\|\partial_3 v\|_0\mbox{ for any }v\in H_0^1,
\end{equation}
thus, by Young's inequality and Korn's inequality \eqref{201901291007}, we see that $\{F(\varpi)\}_{\varpi\in \mathcal{A}}$ has an upper bound for any $\varpi\in \mathcal{A}$.
Hence  there exists  a maximizing sequence $\{w^n\}_{n=1}^\infty\subset \mathcal{A}$, which satisfies $\alpha=\lim_{n\to\infty} F(w_n)$. Moreover, making use of \eqref{201805072235}, the fact $\|\sqrt{ {\rho}}w^n\|_0=1$, trace estimate \eqref{201901281645} and Young's and Korn's inequalities, we have $\|w^n\|_1+\vartheta|\nabla_{\mm{h}}w^n_3|_0 \leqslant c_1$ for some constant $c_1$, which is independent of $n$. Thus, by the well-known Rellich--Kondrachov  compactness theorem and \eqref{201805072235},  there exist a subsequence, still labeled by  $w^n$, and a function $w\in \mathcal{A}$, such that
$$\begin{aligned}
& w^n\rightharpoonup w\mbox{ in }H^1_\sigma,\ w^n\to w\mbox{ in }L^2,\ w^n|_{y_3=0}\to w|_{y_3=0}\mbox{ in }L^2(\mathbb{T}),\\
&  w^n_3|_{y_3=0}\rightharpoonup w_3|_{y_3=0}\mbox{ in }H^1(\mathbb{T})\mbox{ if }\vartheta\neq 0.
\end{aligned}$$
Exploiting the above convergence results, and the lower semicontinuity of weak convergence, we have
\begin{equation*}
-\alpha =\liminf_{n\to \infty}(-F(w^n))
\geqslant -F(w) \geqslant -\alpha.\end{equation*}
Hence $w$ is a maximum point of the functional $F(\varpi)$ with respect to $\varpi\in\mathcal{A}$.

Obviously, $w$ constructed above is also a maximum point of the functional $F(\varpi )/  \|\sqrt{\rho}\varpi\|^2_0$ with respect to  $\varpi\in H_{\sigma, \vartheta}^1$. Moreover $\alpha= F(w)/\|\sqrt{\rho}w\|^2_0$. Thus, for any given $\varphi \in H^1_{\sigma,\vartheta} $, the point $t=0$ is the maximum point of the function
$$I(t):=F(w+t\varphi )-\int \alpha {\rho} |w+t\varphi|^2 \mm{d}y\in C^1(\mathbb{R}).$$
Then, by computing out $I'(0)=0$, we have the weak form:
\begin{align}
\label{201805161108}
& \frac{s}{2}\int \mu \mathbb{D} w:\mathbb{D}\varphi\mm{d}y +\vartheta  \int_\Sigma \nabla_{\mm{h}}w_3\cdot \nabla_{\mm{h}}\varphi_3 \mm{d}y_{\mm{h}}  = g\llbracket\rho\rrbracket \int_\Sigma w_3\varphi_3 \mm{d}y_{\mm{h}}- \alpha \int \rho w\cdot \varphi \mm{d}y.
\end{align}
Noting that \eqref{201805161108} is equivalent to
\begin{align}
&s \int\mu  \mathbb{D} w:\nabla \varphi\mm{d}y+ \vartheta  \int_\Sigma\nabla_{\mm{h}}w_3\cdot \nabla_{\mm{h}}\varphi_3\mm{d}y_{\mm{h}}  = g\llbracket\rho\rrbracket \int_\Sigma w_3\varphi_3\mm{d}y_{\mm{h}}-\alpha  \int \rho w\cdot \varphi\mm{d}y. \nonumber
\end{align}
The means that $w$ is a weak solution of the modified problem \eqref{2016040614130843xd}.
\hfill $\Box$
\end{pf}

\subsection{Improving the regularity of weak solution}

By Proposition \ref{201811241541}, the boundary-value problem \eqref{2016040614130843xd} admits a weak solution $w\in H^1_{\sigma,\vartheta}$. Next we further improve the regularity of $w$.
\begin{pro}
\label{201901281636}
Let $w$ be a weak solution of the boundary-value problem \eqref{2016040614130843xd}. Then
$w\in H^\infty$.
\end{pro}
\begin{pf}
To begin with, we shall establish the following preliminary conclusion:

\emph{For any $i\geqslant 0$, we have}
\begin{equation}
w \in H^{1,i}_{\sigma,\vartheta}
\label{201806181000}
\end{equation}
\emph{and}
\begin{align}
&\frac{1}{2}\int s\mu\mathbb{D} \partial_{\mm{h}}^i w :\mathbb{D}\varphi \mm{d}y + \vartheta\int_\Sigma \nabla_{\mm{h}} \partial_{\mm{h}}^i w_3\cdot\nabla_{\mm{h}}  \varphi_3\mm{d}y_{\mm{h}}\nonumber \\
&  = g\llbracket\rho\rrbracket \int_\Sigma  \partial_{\mm{h}}^i w_3\varphi_3\mm{d}y_{\mm{h}}-\alpha \int \rho   \partial_{\mm{h}}^i w\cdot   \varphi \mm{d}y.
\label{201805161108sdfs}
\end{align}

Obviously, by induction, the above assertion reduces to verify the following recurrence relation:

\emph{For given $i\geqslant 0$, if $w \in H^{1,i}_{\sigma,\vartheta}$ satisfies \eqref{201805161108sdfs}
for any $\varphi \in H^1_{\sigma,\vartheta} $,  then}
\begin{equation}
\label{201806181412}
w \in H^{1,i+1}_{\sigma,\vartheta}
\end{equation}\emph{ and
 $w$ satisfies}
\begin{align}
&\frac{s}{2}\int  \mu \mathbb{D} \partial_{\mm{h}}^{i+1} w :\mathbb{D} \varphi \mm{d}y+ \vartheta \int_\Sigma \nabla_{\mm{h}}  \partial_{\mm{h}}^{i+1} w_3\cdot\nabla_{\mm{h}}  \varphi_3\mm{d}y_{\mm{h}} \nonumber \\
&   =
g\llbracket\rho\rrbracket \int_\Sigma  \partial_{\mm{h}}^{i+1} w_3\varphi_3 \mm{d}y_{\mm{h}}-\alpha  \int \rho \partial_{\mm{h}}^{i+1} w\cdot   \varphi \mm{d}y.\label{201805161108sdfssafas}
\end{align}
Next we verify the above recurrence relation by method of difference quotients.

Now we assume that $w \in H^{1,i}_{\sigma,\vartheta}$ satisfies \eqref{201805161108sdfs}
for any $\varphi \in H^1_{\sigma,\vartheta} $. Noting that $\partial_{\mm{h}}^iw \in H^1_{\sigma,\vartheta}$, we can deduce from \eqref{201805161108sdfs} that, for $j=1$ and $2$,
\begin{align}
& \frac{s}{2}\int \mu  \mathbb{D} \partial_{\mm{h}}^i w : \mathbb{D}  D_j^h \varphi
 \mm{d}y+ \vartheta\int_\Sigma \nabla_{\mm{h}} \partial_{\mm{h}}^i w_3 \cdot \nabla_{\mm{h}}  D_j^h  \varphi_3 \mm{d}y_{\mm{h}} \nonumber \\
& = g\llbracket\rho\rrbracket \int_\Sigma \partial_{\mm{h}}^i w _3  D_j^h  \varphi_3\mm{d}y_{\mm{h}}-\alpha \int \rho
  \partial_{\mm{h}}^i w\cdot  D_j^h \varphi  \mm{d}y \nonumber
\end{align}
and
\begin{align}
& \frac{s}{2}\int \mu \mathbb{D} \partial_{\mm{h}}^i w : \mathbb{D} D_j^{-h} D_j^h \partial_{\mm{h}}^i w
 \mm{d}y+ \vartheta \int_\Sigma \nabla_{\mm{h}}\partial_{\mm{h}}^i w_3\cdot \nabla_{\mm{h}} D_j^{-h} D_j^h\partial_{\mm{h}}^i w_3 \mm{d}y_{\mm{h}}  \nonumber \\
&  = g\llbracket\rho\rrbracket \int_\Sigma   \partial_{\mm{h}}^i w_3 D_j^{-h} D_j^h \partial_{\mm{h}}^i w_3\mm{d}y_{\mm{h}}-\alpha  \int \rho
 \partial_{\mm{h}}^i w\cdot D_j^{-h}  D_j^h  \partial_{\mm{h}}^i w  \mm{d}y, \nonumber
\end{align}
which yield that
\begin{align}
&\frac{s}{2}\int \mu  \mathbb{D} D_j^{-h} \partial_{\mm{h}}^i w : \mathbb{D} \varphi
 \mm{d}y +\vartheta \int_\Sigma \nabla_{\mm{h}} D_j^{-h} \partial_{\mm{h}}^i w_3\cdot \nabla_{\mm{h}}    \varphi_3 \mm{d}y_{\mm{h}}\nonumber \\
& = g\llbracket\rho\rrbracket \int_\Sigma D_j^{-h} \partial_{\mm{h}}^i w_3 \varphi_3\mm{d}y_{\mm{h}}-\alpha \int \rho
  D_j^{-h}\partial_{\mm{h}}^i w\cdot \varphi\mm{d}y ,
\label{20180612713459}
\end{align}
and
\begin{align}
&\|\sqrt{s\mu }\mathbb{D} D_j^h\partial_{\mm{h}}^i w\|^2_0/2+ \vartheta|D_j^h\nabla_{\mm{h}} \partial_{\mm{h}}^i w |^2_0
\nonumber \\
& \lesssim  g\llbracket\rho\rrbracket |D_j^h\partial_{\mm{h}}^i w_3 |^2_0 +|\alpha|\|\sqrt{ \rho}
D_j^h\partial_{\mm{h}}^i w \|^2_0, \label{201806161512}
\end{align}
resp..

By Korn's inequality,
$$\| D_j^h \partial_{\mm{h}}^i w\|^2_{1}\lesssim \|\sqrt{s\mu+\kappa\rho}\mathbb{D} D_j^h \partial_{\mm{h}}^i w\|^2_0,$$
thus, using \eqref{201805072235}, Young's inequality, and the first conclusion in Lemma \ref{xfsddfsf2018050813379safdadf} , we further deduce from \eqref{201806161512} that
\begin{align}
  \|D^h_{\mm{h}}  \partial_{\mm{h}}^i w\|^2_1+ \vartheta| D^h_{\mm{h}}\nabla_{\mm{h}} \partial_{\mm{h}}^i w|^2_0
\lesssim  \|D^h_{\mm{h}} \partial_{\mm{h}}^i w \|^2_0\lesssim \| \nabla_{\mm{h}} \partial_{\mm{h}}^i  w \|^2_0\lesssim 1.\nonumber
\end{align}
Thus, using \eqref{201805072235}, trace estimate \eqref{201901281645} and the second conclusion in Lemma \ref{xfsddfsf2018050813379safdadf},  there exists a subsequence of $\{-h\}_{h\in \mathbb{R}}$ (still denoted by $-h$) such that
\begin{equation}
\label{201806127345}
 \left\{
   \begin{array}{ll}
 D^{-h}_{\mm{h}} \partial_{\mm{h}}^i  w \rightharpoonup \nabla_{\mm{h}}\partial_{\mm{h}}^i w \mbox{ in }H^1_\sigma,\  D^{-h}_{\mm{h}} \partial_{\mm{h}}^i  w \to \nabla_{\mm{h}}\partial_{\mm{h}}^i w \mbox{ in }L^2, &\\
 D^{-h}_{\mm{h}} \partial_{\mm{h}}^i  w|_{y_3=0} \to \nabla_{\mm{h}}\partial_{\mm{h}}^i w |_{y_3=0} \mbox{ in }L^2(\mathbb{T})& \\
 D^{-h}_{\mm{h}}\partial_{\mm{h}}^i w|_{\Sigma}\rightharpoonup \nabla_{\mm{h}}\partial_{\mm{h}}^i w|_{\Sigma}\mbox{ in }H^1(\mathbb{T})\mbox{ if }\vartheta\neq 0.&
   \end{array}
 \right.
\end{equation}
Using  regularity of $w$ in \eqref{201806127345} and the fact $w\in H_{\sigma,\vartheta}^{1,i}$, we have \eqref{201806181412}.
In addition, exploiting the limit results in \eqref{201806127345},  we can deduce \eqref{201805161108sdfssafas} from \eqref{20180612713459}.
This complete the proof of the recurrence relation, and thus \eqref{201806181000} holds.

With \eqref{201806181000} in hand, we can consider a stratified   Stokes problem:
 \begin{equation}\label{2016040614130843x}      \left\{  \begin{array}{ll}
s \nabla \beta^k -s\mu \Delta\omega^k =-\alpha \rho\partial_{\mm{h}}^k w
&\mbox{ in } \Omega,\\
\mm{div}\omega^k=0&\mbox{ in } \Omega,\\
 \llbracket \omega^k  \rrbracket=0,\
\llbracket( s  \beta^k I -  s\mu\mathbb{D}\omega^k)e_3 \rrbracket= \partial_{\mm{h}}^k\mathcal{L}^1&\mbox{ on }\Sigma,
\\ \omega^k =0  &\mbox{ on }\Sigma_{-}^+,\end{array}\right.
\end{equation}
where $k\geqslant 0$ is a given integer, and we have defined that
$$
\begin{aligned}
& \mathcal{L}^1 :=g \llbracket\rho \rrbracket w_3e_3
 + \vartheta\Delta_{\mm{h}}w_3 e_3.
\end{aligned}$$
Recalling the regularity \eqref{201806181000} of $w$, we see that $\partial_{\mm{h}}^kw\in L^2$, and $ \partial_{\mm{h}}^k\mathcal{L}^1 \in H^{1}(\mathbb{T})$. Applying the existence theory of stratified   Stokes problem (see Lemma \ref{xfsddfsf201805072212}), there exists a unique strong solution $(\omega^k,\beta^k)\in H^2\times \underline{H}^1$ of the above problem \eqref{2016040614130843x}.

Multiplying \eqref{2016040614130843x}$_1$ by $\varphi\in H^1_{\sigma,\vartheta}$ in $L^2$ (i.e., taking the inner product in $L^2$), and using the integration by parts and  \eqref{2016040614130843x}$_2$--\eqref{2016040614130843x}$_4$, we have
\begin{align}
 &\frac{s }{2}\int \mu\mathbb{D}\omega^k:\mathbb{D}\varphi \mm{d}y \nonumber
\\
& = g\llbracket\rho\rrbracket \int_\Sigma \partial_{\mm{h}}^k w_3\varphi_3
\mm{d}y_{\mm{h}}  -  \int_\Sigma \vartheta\partial_{\mm{h}}^k \nabla_{\mm{h}}w_3\cdot \nabla_{\mm{h}}\varphi_3\mm{d}y_{\mm{h}} -\int \alpha \rho \partial_{\mm{h}}^k w\varphi\mm{d}y.
\label{201808072057}
\end{align}

Subtracting the two identities  \eqref{201805161108sdfs} and \eqref{201808072057}  yields that
$$s \int \mu\mathbb{D}(\partial_{\mm{h}}^k  w-\omega^k):\mathbb{D} \varphi \mm{d}y=0.
$$
Taking $\varphi:= \partial_{\mm{h}}^k w-\omega^k\in H^1_{\sigma,\vartheta}$ in the above identity, and using the Korn's inequality, we find that $\omega^k= \partial_{\mm{h}}^k w$. Thus we immediately see that
\begin{equation}
\label{201806181507}
\partial_{\mm{h}}^k  w\in H^{2}\mbox{ for any }k\geqslant 0,
\end{equation}
which implies $\partial_{\mm{h}}^kw\in H^1$, and $\partial_{\mm{h}}^k\mathcal{L}^2\in H^{2}(\mathbb{T})$ for any $k\geqslant 0$. Thus, applying the stratified Stokes estimate \eqref{2011805302036} to \eqref{2016040614130843x}, we have
\begin{equation}
\label{201806181507fdsdsgsdfgsg}
\partial_{\mm{h}}^k  w\in H^{3} \mbox{ for any }k\geqslant 0.
\end{equation}
Obviously, by induction, we can easily follow the improving regularity method from \eqref{201806181507} to \eqref{201806181507fdsdsgsdfgsg} to deduce that $w\in H^\infty$.
In addition, we have $\beta:=\beta^0\in H^\infty$; moreover,  $\beta^k$ in \eqref{2016040614130843x}  is equal to $\partial^k_{\mm{h}}\beta$.

Finally, recalling  the embedding $H^{k+2}\hookrightarrow C^0(\overline{\Omega})$ for any $k\geqslant 0$, we easily see that $(w,\beta)$ constructed above is indeed a classical solution to the modified problem \eqref{2016040614130843xd}.
\end{pf}

\subsection{Some properties of the function $\alpha(s)$}

In this subsection, we shall derive some properties of the function $\alpha(s)$, which make sure the existence of fixed point of $\sqrt{\alpha(s)}$ in $\mathbb{R}^+$.
\begin{pro}\label{201901261856}
 For given $\vartheta\in \mathbb{R}^+_0$, we have
\begin{align}
\label{201702081046}
&\alpha(s_2)  <\alpha(s_1)\mbox{ \emph{ for any }}s_2  >s_1>0,\\
\label{201702081047}&
\alpha(s) \in C^{0,1}_{\mm{loc}}(\mathbb{R}^+),\\
&\label{201702081122n}
\alpha(s)>0\mbox{ \emph{ on some interval }}(0,c_2)\mbox{ for } \vartheta\in [0, \vartheta_{\mm{c}}),\\
&\alpha(s)<0\mbox{ \emph{ on some interval }}(c_3,\infty). \label{201702081122}
\end{align}
\end{pro}
\begin{pf}
To being with, we  verify \eqref{201702081046}.
For given $s_2>s_1$, then there exist  $v^{s_2}\in \mathcal{A}$ such that
$\alpha(s_2)  = F(v^{s_2},s_2)$.
Thus, by Korn's inequality and the fact $\|\sqrt{\rho}v^{s_2}\|_0=1$,
$$
 \alpha(s_1)\geqslant F(v^{s_2},s_1) = \alpha(s_2) +
 ( s_2- {s_1}) \|\sqrt{\mu }\mathbb{D}v^{s_2}\|_0^2/2 > \alpha(s_2),
$$
 which yields \eqref{201702081046}.

Now we turn to prove \eqref{201702081047}. Choosing a bounded interval $[c_4,c_5]\subset (0,\infty)$, then, for any $s\in [c_4,c_5] $, there exists a function $v^s$ satisfying $\alpha(s)=F(v^{s},s)$. Thus, by the monotonicity \eqref{201702081046}, we have
 $$\alpha(c_5)+
c_4\|\sqrt{\mu }\mathbb{D}v^s\|_0^2/4\leqslant F(v^{s},s/2) \leqslant \alpha(s/2) \leqslant \alpha( c_4/2),$$
  which yields
$$\|\sqrt{\mu } \mathbb{D}v^s\|_0^2/2 \leqslant  2(\alpha(c_4/2)-\alpha(c_5))/c_4=:\xi\mbox{ for any }s\in [c_4,c_5].$$
Thus, for any $s_1$, $s_2\in [c_4,c_5]$,
$$
\begin{aligned}
\alpha(s_1)-\alpha(s_2)\leqslant & F(v^{s_1},s_1)-F(v^{s_1},s_2)\leqslant
\xi|  {s_2}-s_1|
\end{aligned}$$
 and $$\alpha(s_2)-\alpha(s_1)\leqslant
\xi|  {s_2}-s_1|,$$
which immediately imply $|\alpha(s_1)-\alpha(s_2)|\leqslant \xi| {s_2}-s_1|$. Hence \eqref{201702081047} holds.

Finally, \eqref{201702081122n}
can be  deduced from  the definition of $\alpha$ by using  Korn's inequality and \eqref{201805072235}, while \eqref{201702081122} is obvious by the definition of $\alpha$, Lemma \ref{201811241500} and \eqref{201811241957}.
\hfill $\Box$
\end{pf}

\subsection{Construction of an interval for fixed point}
Let
$\mathfrak{I}:=\sup\{$all the real constant $s$, which satisfy that $\alpha( \tau)>0$ for any $\tau\in (0,s)\}$.
By virtue of \eqref{201702081122n} and \eqref{201702081122}, $ \mathfrak{I}\in \mathbb{R}_+$. Moreover, $\alpha(s)>0$ for any $s\in (0,\mathfrak{I})$, and, by the continuity of $\alpha(s)$,
\begin{equation}\label{nzerolin}
 \alpha( \mathfrak{I})=0.
\end{equation}
Using the monotonicity and the upper boundedness of $\alpha(s)$, we see that
 \begin{equation}\label{zeron}
 \lim_{s\rightarrow 0}\alpha(s)=\varsigma\mbox{ for some postive constant }\varsigma.
 \end{equation}

Now, exploiting \eqref{nzerolin},  \eqref{zeron} and the continuity of ${\alpha}(s)$ on $ (0,\mathfrak{I})$,
we find by a fixed-point argument on $(0,\mathfrak{I})$ that there is a unique $\Lambda\in(0,\mathfrak{I})$ satisfying
\begin{equation}
\label{201901291010}
  \Lambda=\sqrt{ \alpha(\Lambda)}=
\sqrt{\sup_{\varpi\in\mathcal{A}}F(\varpi, \Lambda )}\in (0,\mathfrak{I}).
\end{equation}
Thus we get a classical solution $(w,\beta) \in H^\infty$ to the boundary problem  \eqref{201604061413} with $\Lambda$ constructed by \eqref{201901291010}.
Moreover,
\begin{equation}
\label{growthnn} \Lambda= \sqrt{ F(w, \Lambda )}>0.
\end{equation}
In addition, \eqref{201602081445MH}  directly follows \eqref{growthnn} and  the fact
$w\in H_\sigma^1$.

\subsection{Largest growth rate}\label{201807111049}

Next we shall prove that $\Lambda$ constructed in previous section is the largest  growth rate of RT instability in the linearized  RT problem, and thus complete the proof of Theorem \ref{201806012301xxxxxxxx}.

\begin{pro}\label{bestgrowth}
Under the assumptions of Theorem \ref{201806012301xxxxxxxx},  $\Lambda>0$  constructed by \eqref{201901291010}  is the largest  growth rate of RT instability in the linearized  RT problem.
\end{pro}
\begin{pf}
Recalling the definition of largest  growth rate, it  suffices to prove that $\Lambda$ enjoys the first  condition in Definition \ref{201804072001}.

Let $u$ be strong solution to the linearized RT problem. Then we derive that,
for a.e. $t\in I_T$ and all $w\in H_{\sigma}^1$,
\begin{align}
\int \rho u_t\cdot w\mm{d}y= &\int (\mu \Delta u-\nabla q)\cdot w\mm{d}y \nonumber \\
=&\int_\Sigma(g\llbracket\rho\rrbracket \eta_3w_3+\vartheta \Delta_{\mm{h}}\eta_3  w_3 )\mm{d}y_{\mm{h}}-\int \mu \mathbb{D}u: \nabla w\mm{d}y.
\end{align}
thus,
\begin{align}
\frac{\mm{d}}{\mm{d}t}
\int \rho u_t\cdot w\mm{d}y
=\int_\Sigma(g\llbracket\rho\rrbracket u_3w_3+\vartheta \Delta_{\mm{h}}u_3w_3 )\mm{d}y_{\mm{h}}-\int \mu \mathbb{D}u_t: \nabla  w\mm{d}y. \label{201901251532}
\end{align}
Using regularity of $(\eta,u)$, we can show that the right hand side of \eqref{201901251532} is bounded above by $A(t)(\|w\|_1+|w|_1)$ for some positive function $A(t)\in L^2(I_T)$.
Then there exists a $f\in L^2(I_T, H^{-1}_{\sigma})$ such that, for a.e. $t\in I_T$,
\begin{equation}
\label{201901291420}
 <f, w>_{H^{-1}_{\sigma}\times H^1_{\sigma}}:= \int_\Sigma(g\llbracket\rho\rrbracket u_3w_3+\vartheta \Delta_{\mm{h}}u_3 w_3 )\mm{d}y_{\mm{h}}-\int \mu \mathbb{D}u_t: \nabla  w\mm{d}y.
 \end{equation}
Hence it follows from  Lemma  \ref{201901251652}  that
$$(\rho u_t)_t=f\in L^2(I_T,H^{-1}_{\sigma}). $$

In addition, by a classical regularization method (referring to Theorem 3 in Chapter 5.9 in \cite{ELGP} and Lemma 6.5 in \cite{NASII04}), we have
\begin{align}
&\frac{1}{2}\frac{\mm{d}}{\mm{d}t}\int \rho|u_t|^2\mm{d}y=2  <\partial_t(\rho u_t), u_t>_{H^{-1}_{\sigma}\times H^1_{\sigma}},\nonumber \\
& \int_\Sigma  \Delta_{\mm{h}}u_3 \partial_t u_3\mm{d}y_{\mm{h}}=\frac{1}{2} \frac{\mm{d}}{\mm{d}t}\int_\Sigma  |\nabla_{\mm{h}}u_3|^2\mm{d}y_{\mm{h}}. \nonumber
\end{align}
Therefore, we can derive from \eqref{201901291420} and the above two identities that
\begin{align}
 \frac{\mm{d}}{\mm{d}t}\left(\|\sqrt{\rho} u_t \|^2_{0}+E(u )\right)+ \|\sqrt{\mu} \mathbb{D}  u _t\|_0^2\mm{d}\tau  =0. \nonumber
\end{align}
Then, integrating the above identity in time from $0$ to $t$ yields that
\begin{align}
 \|\sqrt{\rho} u_t \|^2_{0}+E(u )+ \int_0^t\|\sqrt{\mu} \mathbb{D}  u _s\|_0^2\mm{d}s  = I^0:= {E}(u |_{t=0})+ \| \sqrt{\rho}u_t\big|_{t=0} \|^2_0. \label{0314}
\end{align}

Using Newton--Leibniz's formula and Young's inequality, we find that
\begin{align}
\Lambda \|\sqrt{\mu}\mathbb{D} u(t)\|_{0}^2
& =\Lambda   \| \sqrt{\mu} \mathbb{D} u_0\|_{0}^2 + 2\Lambda\int_0^t\int  \mu
\mathbb{D} u(s):\mathbb{D} u_s  \mm{d} y\mathrm{d}s\nonumber  \\
& \leqslant \Lambda  \|\sqrt{\mu} \mathbb{D}  u_0\|_{0}^2 +\int_0^t  \|\sqrt{\mu} \mathbb{D} u_s \|_{0}^2
\mathrm{d}s +\Lambda^2\int_0^t  \|\sqrt{\mu} \mathbb{D}  u (s) \|_{0}^2 \mathrm{d}s. \label{0316xx}
\end{align}
In addition, by \eqref{201901251824}, we have
  \begin{equation}\label{0302xx}
  -\mathcal{E}(u) \leqslant {\Lambda^2}\|  \sqrt{\rho} {u} \|^2_0
 +\frac{\Lambda}{2}  \|\sqrt{\mu }  \mathbb{D}   u\|^2_0.
\end{equation}
Thus, we infer from \eqref{0314}--\eqref{0302xx} that
\begin{align}
&\frac{1}{\Lambda}\|\sqrt{ {\rho}}  u_t  \|^2_{0}+
 \frac{1}{2} \|\sqrt{\mu}\mathbb{D} u (t)\|_{0}^2 \nonumber  \\
& \leqslant   {\Lambda} \|\sqrt{ {\rho}} u (t)\|^2_{0}+  {\Lambda}\int_0^t  \|\sqrt{\mu}
\mathbb{D} u(s)  \|_{0}^2 \mm{d}s +\frac{I^0+ \Lambda   \|\sqrt{\mu}\mathbb{D}  u_0\|_{0}^2 }{\Lambda}. \label{inequalemee}
\end{align}
Recalling that
\begin{equation*}\begin{aligned}
\Lambda\frac{\mm{d}}{\mm{d}t}\|\sqrt{ {\rho}} u  \|^2_{0}
= 2\Lambda\int\rho u(t)\cdot u_t\mm{d} y \leqslant \|\sqrt{\rho} u_t \|^2_{0} +\Lambda^2\|\sqrt{ {\rho}} u(t)\|^2_{0},
\end{aligned}\end{equation*}
we further deduce from \eqref{inequalemee} the differential inequality:
\begin{equation*}
\begin{aligned}
& \frac{\mm{d}}{\mm{d}t}\|\sqrt{ {\rho}} u  \|^2_{0}+ \frac{1}{2}\| \sqrt{\mu} \mathbb{D}  u(t)\|_{0}^2  \\
& \leqslant  2\Lambda\left( \|\sqrt{\rho} u(t)\|^2_{0} +\frac{1}{2}\int_0^t \|\sqrt{\mu} \mathbb{D} u(s) \|_{0}^2
\mathrm{d}s\right) +\frac{I^0 + \Lambda \|\sqrt{\mu} \mathbb{D} u_0\|_{0}^2 }{\Lambda}.
\end{aligned}
\end{equation*}

 Applying Gronwall's inequality \cite[Lemma 1.2]{NASII04} to the above inequality, one concludes
\begin{align}
  \|\sqrt{ {\rho}} u (t)\|^2_{0}+\frac{1}{2}\int_0^t  \|\sqrt{\mu}  \mathbb{D} u(s)  \|^2_{0} \mm{d}s
 \leqslant \left(\|\sqrt{ {\rho}}u^0\|_{0}^2+\frac{ I^0+ \Lambda   \|\sqrt{\mu}  \mathbb{D}  u^0\|_{0}^2 }{2\Lambda^{2}}\right) e^{2\Lambda t},\label{201901291440}
 \end{align}
which, together with \eqref{inequalemee}, yields
  \begin{align}
\frac{1}{\Lambda}\|\sqrt{ {\rho}}  u_t (t)\|^2_{0} +
 \frac{1}{2} \|\sqrt{\mu}  \mathbb{D}  u (t)\|_{0}^2 \leqslant & 2\left(\Lambda\|\sqrt{ {\rho}}u^0\|_{0}^2+\frac{ I^0+ \Lambda   \|\sqrt{\mu}
  \mathbb{D}  u^0\|_{0}^2  }{2\Lambda }\right) e^{2\Lambda t}\nonumber \\
  & +\frac{I^0+ \Lambda  \|\sqrt{\mu}  \mathbb{D}  u^0\|_{0}^2 }{\Lambda}.\label{201901291443}
\end{align}

Multiplying  \eqref{201811242002xx}$_2$ by $u_t$ in $L^2$ and using the integral by parts, we get
\begin{align}
 \int \rho| u_{t}|^2\mm{d}y=\int_{\Sigma} \llbracket   q \rrbracket \partial_tu_3 \mm{d}y_{\mm{h}}
 -\int \mu \Delta u\cdot u_t\mm{d}y.
\label{201806271923}
\end{align}
Exploiting  \eqref{06201929201811041930}, we can estimate that
\begin{align}
\int_{\Sigma} \llbracket   q \rrbracket \partial_tu_3 \mm{d}y_{\mm{h}}
\lesssim &  |\llbracket   q \rrbracket |_{1/2}|\partial_t u_3|_{-1/2}\lesssim |\llbracket   q \rrbracket |_{1/2}\|u_t\|_{0}.
 \nonumber
\end{align}
In addition, using \eqref{201811242002xx}$_5$ and trace estimate \eqref{201901281645},  we have
$$|\llbracket   q \rrbracket |_{1/2}\lesssim \|\eta\|_3+\|u\|_2.$$
Using the above two estimates, we can derive from \eqref{201806271923} that
\begin{align}
\| u_{t}\|^2_0\lesssim \|\eta\|_3+\|u\|_2 ,\nonumber
\end{align}
which implies that
$$ \| \sqrt{\rho}u_t\big|_{t=0} \|^2_0 \lesssim \|(\eta^0,u^0)\|_3.$$
By the above estimate and Korn's inequality, we derive from \eqref{201901291440} and  \eqref{201901291443}  that
\begin{align}
\| u \|_{1}^2+\| u_t \|^2_{0 }+ \int_0^t\| u(s) \|^2_{1}\mm{d}s\lesssim
  e^{2\Lambda t}(\|\eta^0\|_3^2+\|u^0\|_2^2) .\nonumber
\end{align}

Finally, from \eqref{201811242002xx}$_1$ we get
\begin{align}
\|\eta(t)\|_{1} \lesssim & \| \eta^0\|_{1} +\int_0^t \| \eta_s\|_{1}  \mm{d}s
\lesssim \|\eta^0\|_{1} + \int_0^t\| u(s)\|_{1}\mm{d}s  \nonumber \\
\lesssim & e^{\Lambda t}(\|\eta^0\|_{3}+\|u^0\|_2).\nonumber
\end{align}
By the two estimates  above, we see that $\Lambda$ satisfies the first  condition in Definition \ref{201804072001}.
The proof is complete.
 \hfill$\Box$
\end{pf}

\section{Effect of surface tension}\label{201901292122}

\subsection{Properties of $\alpha(s,\vartheta)$ with respect to $\vartheta$}

   To emphasize the dependence of  $\Lambda$ and $\mathcal{G}$ upon $\vartheta$, we will denote them by $\Lambda_\vartheta$ and $\mathcal{G}_\vartheta$, respectively.
To prove  Theorem \ref{201901281046}, we shall
further derive the relations \eqref{201901251413} and \eqref{201901252005} of surface tension coefficient and the largest growth rate. To this end, we need the following auxiliary conclusions:
\begin{pro}\label{lem:0201}
Let  $g>0$, $\rho>0$ and $\mu>0$ are given.
\begin{enumerate}[(1)]
\item Strict monotonicity: if $\vartheta_{1}$ and $\vartheta_{2}$ are constants satisfying $0\leqslant \vartheta_{1}<\vartheta_{2}$, then
\begin{equation}\label{0217}
\alpha(s,\vartheta_{2})<\alpha(s,\vartheta_{1}).
\end{equation}
for any given $s>0$. Moreover, if  $\vartheta_2$ further satisfies $\vartheta_2<\vartheta_{\mm{c}}$,
\begin{equation}\label{0218}
\mathcal{G}_{\vartheta_{\mm 1}}>\mathcal{G}_{\vartheta_{\mm 2}},
\end{equation}
where
\begin{equation}\label{0219}
\mathcal{G}_{\vartheta_i}:=\sup\{s\in\mathbb{R}~|~\alpha(\tau,\vartheta_i)>0\mbox{ for any}\;\;\tau\in(0,s)\}\mbox{ and }\alpha(\mathcal{G}_{\vartheta_i},\vartheta_i)=0.
\end{equation}
\item Continuity: for given $s>0$, $\alpha(s,\vartheta)\in C^{0,1}_{\mm{loc}}(\mathbb{R}_+)$
 with respect to the variable $\vartheta$.
\end{enumerate}
\end{pro}
\begin{pf}
(1) Let $s>0$ be fixed, and $0\leqslant\vartheta_{1}<\vartheta_{2}$. Then there exist functions $ {w}^{\vartheta_i}\in H^\infty\cap \mathcal{A}_{\vartheta_i}$, $i=1,2$, such that
\begin{equation*}\begin{aligned}
\alpha(s,\vartheta_i)= {E}
( {w}
^{\vartheta_i})-\vartheta|\nabla_{\mm{h}} {w}_3^{\vartheta_i}|_{0}^2.
\end{aligned}
\end{equation*}
where $ {E}( {w}^{\vartheta_i}):=g\llbracket\rho
\rrbracket| {w}^{\vartheta_i}_{3}|_0^{2}
-{s}\|\sqrt{\mu} \mathbb{D} {w}^{\vartheta_i}\|_0^2/2$. Since $ {w}^{\vartheta_i}\in\mathcal{A}_{\vartheta_i}$, by virtue of \eqref{201602081445MH}  and \eqref{201806101508}, we have
\begin{equation*}\begin{aligned}
0<| {w}^{\vartheta_2}|_{0}\lesssim | \nabla_{\mm{h}} {w}^{\vartheta_2}|_0^2,
\end{aligned}
\end{equation*}
and, thus
\begin{equation*}\begin{aligned}
\alpha(s,\vartheta_{2})\leqslant \alpha(s,\vartheta_{1})+ {(\vartheta_{1}-\vartheta_{2})
}   |\nabla_{\mm{h}} {w}_3^{\vartheta_2}|_{0}^2<\alpha(s,\vartheta_{1}).
\end{aligned}
\end{equation*}
This yields the desired conclusion \eqref{0217}.

Next we prove  \eqref{0218} by contradiction. If $\mathcal{G}_{\vartheta_{\mm1}}<\mathcal{G}_{\vartheta_{\mm2}}$, then we get from  \eqref{0217} and the strict monotonicity of $\alpha(s,\cdot)$ with respect to $s$ that
\begin{equation*}\begin{aligned}
0=\alpha(\mathcal{G}_{\vartheta_{\mm2}},\vartheta_{2})
<\alpha(\mathcal{G}_{\vartheta_{\mm2}},\vartheta_{1})
<\alpha(\mathcal{G}_{\vartheta_{\mm1}},\vartheta_{1})=0,
\end{aligned}
\end{equation*}
which is a paradox. If $\mathcal{G}_{\vartheta_{\mm1}}=\mathcal{G}_{\vartheta_{\mm2}}$, exploiting \eqref{0217}, we have
\begin{equation*}\begin{aligned}
0=\alpha(\mathcal{G}_{\vartheta_{\mm2}},\vartheta_{2})<
\alpha(\mathcal{G}_{\vartheta_{\mm2}},\vartheta_{1})=
\alpha(\mathcal{G}_{\vartheta_{\mm1}},\vartheta_{1})=0,
\end{aligned}
\end{equation*}
which is also a paradox. Thus we immediately get the desired conclusion.

(2) Let $s>0$ be fixed. We choose a bounded interval  $[b_1,b_2]\subset \mathbb{R}_+$. Then, for any given $\theta\in [ b_1/2,b_2]$, there is a function $ {w}^\theta\in\mathcal{A}_{\vartheta}$ satisfying $\alpha(s,\theta)={E}( {w}^\theta)
-\theta|\nabla_{\mm{h}} {w}_3^{\theta}|_{0}^2$. Thus, in view of  the monotonicity of $\alpha(\cdot,\theta)$, we know that
\begin{align}
\alpha(s,b_2)+b_1|\nabla_{\mm{h}} {w}_3^{\theta}|_{0}^2/2\leqslant &    \alpha(s,\theta)+\theta|\nabla_{\mm{h}} {w}_3^{\theta }|_{0}^2/2 \nonumber \\
= & \alpha(s,\theta/2 ) \leqslant \alpha(s,b_1/2),
\end{align}
which yields
$$
\begin{aligned}
|\nabla_{\mm{h}} {w}_3^{\vartheta}|_{0}^2\leqslant  2 (\alpha(s,b_1/2)-\alpha(s,b_2)
 )/b_1 :=K(s)
\;\;\mbox{for any}\;\vartheta\in(b_1,b_2).
\end{aligned}$$
Thus, for any $\vartheta_1$, $\vartheta_2\in [b_1,b_2]$,
$$
\begin{aligned}
\alpha(s,\vartheta_1)-\alpha(s,\vartheta_2)\leqslant & {E}( {w}^{\vartheta_1})-\vartheta_1 |\nabla_{\mm{h}} {w}_3^{\theta_1 }|_{0}^2
-\big({E}( {w}^{\vartheta_1})-\vartheta_2|\nabla_{\mm{h}} {w}_3^{\theta_1 }|_{0}^2)\big)\nonumber \\
\leqslant & K(s)|\vartheta_2-\vartheta_1|.
\end{aligned}$$
Reversing the role of the indices $1$ and $2$ in the derivation of the above inequality, we
obtain the same boundedness with the indices switched. Therefore, we deduce that
\begin{equation*}\begin{aligned}
|\alpha(s,\vartheta_{1})-\alpha(s,\vartheta_{2})|\leqslant K(s)|\vartheta_{1}-\vartheta_{2}|,
\end{aligned}
\end{equation*}
which yields $\alpha(s,\vartheta)\in C^{0,1}_{\mm{loc}}(\mathbb{R}_+)$.
This completes the proof. \hfill $\Box$
\end{pf}

\subsection{Proof of  Theorem \ref{201901281046}}\label{201901261854}

\emph{First, we verify the monotonicity of $\Lambda_{\vartheta}$ with respect to the variable $\vartheta\in [0, \vartheta_{\mm{c}})$.}

For given two constants $\vartheta_{1}$ and $\vartheta_{2}$ satisfying $0\leqslant \vartheta_{1}<\vartheta_{2}< \vartheta_{\mm{c}}$, then there exist two associated curve functions $\alpha(s,\vartheta_{1})$ and $\alpha(s,\vartheta_{2})$ defined in $(0,\vartheta_{\mm{c}})$. By the first assertion in Proposition \ref{lem:0201}.
\begin{equation*}
\alpha(s,\vartheta_{1})>\alpha(s,\vartheta_{2}).
\end{equation*}
On the one hand, the fixed-point $\Lambda_{\vartheta_{i}}$ satisfying $\Lambda_{\vartheta_{i}}=\sqrt{\alpha(\Lambda_{\vartheta_{i}})}$ can be obtained from the intersection point of the two curves $y=s$ and $y=\sqrt{\alpha(s,\vartheta_{i})}$ on $(0,\mathcal{G}_{\vartheta_{i}})$ for $i=1$ and $2$. Thus we can immediately observe the monotonicity
\begin{equation}\label{0224}\begin{aligned}
\Lambda_{\vartheta_{1}}>\Lambda_{\vartheta_{2}}\mbox{ for }0\leqslant \vartheta_{1}<\vartheta_{2}< \vartheta_{\mm{c}}.
\end{aligned}
\end{equation}

\emph{Second, we prove the continuity for $\Lambda_{\vartheta}$. }

We choose a constant $\vartheta_{0}>0$ and an associated function $\alpha(s,\vartheta_{0})$. Noting that $\alpha(\Lambda_{\vartheta_{0}},\vartheta_{0})=\Lambda^{2}_{\vartheta_{0}}>0$ and $\alpha(\cdot,\vartheta )\in C^{0,1}_{\mm{loc}}[0,\vartheta_{\mm{c}})$ are strictly decreasing and continuous with respect to $\vartheta$, then, for any given $\varepsilon>0$, there exists a constant $\delta>0$, such that
\begin{equation*}\begin{aligned}
  (\vartheta_{0}-\delta,\vartheta_{0}+\delta) \subset (0,\vartheta_{\mm{c}} ),\ \alpha(\Lambda_{\vartheta_{0}},\vartheta_{0}+\delta)>0,\ 0<\sqrt{\alpha(\Lambda_{\vartheta_{0}},\vartheta_{0})}-\sqrt{\alpha(\Lambda_{\vartheta_{0}},\vartheta_{0}+\delta)}<\varepsilon
\end{aligned}
\end{equation*}
and
\begin{equation*}\begin{aligned} 0<\sqrt{\alpha(\Lambda_{\vartheta_{0}},\vartheta_{0}-\delta)}-\sqrt{\alpha(\Lambda_{\vartheta_{0}},\vartheta_{0})}<\varepsilon.
\end{aligned}
\end{equation*}
In particular, we have
\begin{equation*}\begin{aligned}
\Lambda_{\vartheta_{0}}-\varepsilon<\sqrt{\alpha(\Lambda_{\vartheta_{0}},\vartheta_{0}+\delta)}\mbox{ and }\sqrt{\alpha(\Lambda_{\vartheta_{0}},\vartheta_{0}-\delta)}<\Lambda_{\vartheta_{0}}+\varepsilon.
\end{aligned}
\end{equation*}
By the monotonicity of $\Lambda_{\vartheta}$ with respect to $\vartheta$, we get
\begin{equation*}\begin{aligned}
\Lambda_{\vartheta_{0}-\delta}>\Lambda_{\vartheta_{0}}>\Lambda_{\vartheta_{0}+\delta}.
\end{aligned}
\end{equation*}
Thus, using the monotonicity of $\alpha(s,\cdot)$ with respect to $s$, we obtain
\begin{equation*}\begin{aligned}
\sqrt{\alpha(\Lambda_{\vartheta_{0}},\vartheta_{0}+\delta)}<\sqrt{\alpha(\Lambda_{\vartheta_{0}+\delta},\vartheta_{0}+\delta)}=\Lambda_{\vartheta_{0}+\delta}
\end{aligned}
\end{equation*}
and
\begin{equation*}\begin{aligned}
\sqrt{\alpha(\Lambda_{\vartheta_{0}},\vartheta_{0}-\delta)}>\sqrt{\alpha(\Lambda_{\vartheta_{0}-\delta},\vartheta_{0}-\delta)}=\Lambda_{\vartheta_{0}-\delta}.
\end{aligned}
\end{equation*}
Chaining the five inequalities above, we immediately get
\begin{equation*}\begin{aligned}
\Lambda_{\vartheta_{0}}-\varepsilon<\Lambda_{\vartheta_{0}+\delta}<\Lambda_{\vartheta_{0}-\delta}<\Lambda_{\vartheta_{0}}+\varepsilon.
\end{aligned}
\end{equation*}
Then, for any $\vartheta\in(\vartheta_{0}-\delta,\vartheta_{0}+\delta)$, we arrive at
$\Lambda_{\vartheta_{0}}-\varepsilon<\Lambda_{\vartheta}<\Lambda_{\vartheta_{0}}+\varepsilon$.
Hence
\begin{equation}\label{0225xx}\begin{aligned}
\Lambda_{\vartheta}\mbox{ is continuous function of }
\vartheta\in (0,\vartheta_{\mm{c}}).
\end{aligned}
\end{equation}

Now  we study the limit of $\Lambda_\vartheta$ as $\vartheta\to 0$.
For any $\varepsilon>0$, there exits a $ {w}\in \mathcal{A}_0 $ such that
\begin{equation}
\label{201901292254}
w_3\neq 0 \mbox{ on }\Sigma\mbox{ and } \Lambda_0 -\varepsilon<   \sqrt{  g\llbracket\rho\rrbracket
| {w}_{3}|_0^{2}
- \Lambda_0\|\sqrt{\mu} \mathbb{D} {w} \|_0^2/2
   }=\Lambda_0.
 \end{equation}
    In addition,
  \begin{equation}
  \Lambda_\vartheta< \Lambda_0. \label{201901292254xx}
  \end{equation}
Thus, making use  of \eqref{201901251824}, \eqref{201901292254}  and \eqref{201901292254xx}, there exists a sufficiently small constant $\vartheta_1\in (0,\vartheta_{\mm{c}})$ such that, for any $\vartheta\in (0,\vartheta_1)$,
\be
\label{lamdfvear}
\Lambda_0 -\varepsilon< \sqrt{  g\llbracket\rho\rrbracket
| {w}_{3}|_0^{2}
- \Lambda_\vartheta\|\sqrt{\mu} \mathbb{D} {w} \|_0^2/2-\vartheta |\partial_{\mm{h}}w_3|^2_0} \leqslant \Lambda_\vartheta< \Lambda_0.
\ee
Hence we get
\begin{equation*}
\lim_{\vartheta\rightarrow 0}\Lambda_\vartheta = \Lambda_0
\end{equation*}
which, together with \eqref{0225xx}, yields that
\begin{equation}\label{0225}\begin{aligned}
\Lambda_{\vartheta}\mbox{ is continuous function of }
\vartheta\in [0,\vartheta_{\mm{c}}).
\end{aligned}
\end{equation}

\emph{Finally, we derive the upper bound \eqref{201901251413} for $\Lambda_\vartheta$.}

 Recalling the definition of $\vartheta_{\mm c}$, we see from \eqref{2018060102215680} that
\begin{equation*}\begin{aligned}
g\llbracket\rho\rrbracket| {w}_{3}|_0^{2}\leqslant  \vartheta_{\mm c} |\nabla_{\mm{h}} {w}_3 |_{0}^2\quad\mbox{for any}\; {w}\in H^1_{\sigma,\Sigma}.
\end{aligned}
\end{equation*}
 Hence, by virtue of \eqref{201901251824}, for any given $\vartheta\in [0,\vartheta_{\mm{c}})$, there exists a $ {w}^{\vartheta}\in\mathcal{A}_\vartheta$ such that
 \begin{align}
0 \leqslant \Lambda_\vartheta^2=
F(w^{\vartheta},\Lambda_\vartheta)\leqslant g\llbracket\rho\rrbracket (\vartheta_{\mm{c}}-\vartheta) | {w}^{\vartheta}_{3}|_0^{2}/\vartheta_{\mm{c}}-\frac{\Lambda_\vartheta}{2}\|\sqrt{\mu }\mathbb{D}  {w}^{\vartheta} \|_0^2,\nonumber
\end{align}
which yields that
 \begin{align}
\Lambda_\vartheta^2+ \frac{\Lambda_\vartheta}{2}\|\sqrt{\mu }\mathbb{D}  {w}^{\vartheta} \|_0^2 \leqslant g\llbracket\rho\rrbracket (\vartheta_{\mm{c}}-\vartheta) | {w}^{\vartheta}_{3}|_0^{2}/\vartheta_{\mm{c}}. \label{201811242131}
\end{align}

By \eqref{201901272014} and trace estimate \eqref{201901281645},
we can  estimate that
$$
\begin{aligned}
 | {w}^{\vartheta}_{3}|_0^{2} \leqslant
 \frac{h_+}{8  \mu_+}\| \sqrt{\mu} \mathbb{D} {w}^{\vartheta}\|_0^2 .
\end{aligned}
 $$
 Similarly, we also have
 $$
\begin{aligned}
 | {w}^{\vartheta}_{3}|_0^{2}
 \leqslant
 \frac{ h_-}{8 \mu_-}\| \sqrt{\mu} \mathbb{D} {w}^{\vartheta}\|_0^2 .
\end{aligned}
 $$
By the above two estimates, we derive from \eqref{201811242131} that
$$
\begin{aligned}
\Lambda_\vartheta \| \sqrt{\mu} \mathbb{D} {w}^{\vartheta}\|_0^2 \leqslant  \frac{ g\llbracket\rho\rrbracket (\vartheta_{\mm{c}}-\vartheta)}{ 4\vartheta_{\mm{c}} }\min \left\{\frac{h_+}{\mu_+},\frac{h_-}{\mu_-}\right\}  \| \sqrt{\mu} \mathbb{D} {w}^{\vartheta}\|_0^2,
\end{aligned}
 $$
 which yields that
\begin{equation}\label{0230xxxx}\begin{aligned}
\Lambda_{\vartheta}\leqslant  \frac{  (\vartheta_{\mm{c}}-\vartheta)}{ 4\max\{L_1^2,L_2^2\} }\min \left\{\frac{h_+}{\mu_+},\frac{h_-}{\mu_-}\right\}.
\end{aligned}
\end{equation}

Noting that $\|\sqrt{\rho}{w}^{\vartheta} \|_0=1$, then, by \eqref{201901301148},
$$
\begin{aligned}
 | {w}^{\vartheta}_{3}|_0^{2} \leqslant
 \frac{2}{\sqrt{\rho_-}}\|\sqrt{\rho_-} {w}^{\vartheta}_3 \|_{L^2(\Omega_-)}\| \partial_3{w}^{\vartheta}_3 \|_{L^2(\Omega_-)}\leqslant
   \frac{\|\sqrt{\mu}\mathbb{D}{w}^{\vartheta}\|_{0}}{\sqrt{2\rho_-\mu_-}}.
\end{aligned}
 $$
 Putting the above estimate into \eqref{201811242131}, and then using Young's inequality, we get
  \begin{align}
\Lambda_\vartheta^2+ \frac{\Lambda_\vartheta}{4}\|\sqrt{\mu }\mathbb{D}  {w}^{\vartheta} \|_0^2 \leqslant \frac{4(g\llbracket\rho\rrbracket (\vartheta_{\mm{c}}-\vartheta))^2 }{\vartheta_{\mm{c}}^2 \rho_-\mu_- \Lambda_\vartheta }, \nonumber \end{align}
which yields that
 \begin{align}
\Lambda_\vartheta^3  \leqslant \frac{4(g\llbracket\rho\rrbracket (\vartheta_{\mm{c}}-\vartheta))^2 }{\vartheta_{\mm{c}}^2 \rho_-\mu_- }.
\end{align}
Similarly, we also have
 \begin{align}
\Lambda_\vartheta^3  \leqslant \frac{4(g\llbracket\rho\rrbracket (\vartheta_{\mm{c}}-\vartheta))^2 }{\vartheta_{\mm{c}}^2 \rho_+\mu_+ }.
\end{align}
Summing up the above two estimates yields that
 \begin{align}
\Lambda_\vartheta  \leqslant
  \left(\frac{4(g\llbracket\rho\rrbracket (\vartheta_{\mm{c}}-\vartheta))^2 }{\vartheta_{\mm{c}}^2 \max\{\rho_+\mu_+ ,
\rho_-\mu_-\}}\right)^{\frac{1}{3}},
\end{align}
which, together with \eqref{0230xxxx}, implies that
 \begin{align}\label{0230}
 \Lambda_\vartheta  \leqslant m. \end{align}
Consequently we complete the proof of  Theorem \ref{201901281046} from \eqref{0224}, \eqref{0225} and \eqref{0230}.

\vspace{4mm} \noindent\textbf{Acknowledgements.}
 The research of Fei Jiang was supported by NSFC (Grant No. 11671086).

\renewcommand\refname{References}
\renewenvironment{thebibliography}[1]{%
\section*{\refname}
\list{{\arabic{enumi}}}{\def\makelabel##1{\hss{##1}}\topsep=0mm
\parsep=0mm
\partopsep=0mm\itemsep=0mm
\labelsep=1ex\itemindent=0mm
\settowidth\labelwidth{\small[#1]}%
\leftmargin\labelwidth \advance\leftmargin\labelsep
\advance\leftmargin -\itemindent
\usecounter{enumi}}\small
\def\newblock{\ }
\sloppy\clubpenalty4000\widowpenalty4000
\sfcode`\.=1000\relax}{\endlist}
\bibliographystyle{model1b-num-names}

\end{document}